\documentclass[twocolumn,apj,numberedappendix,twocolappendix]{openjournal}

\usepackage{xcolor}
\definecolor{linkcolor}{rgb}{0.0,0.3,0.5}
\usepackage{textgreek}

\usepackage{graphicx}
\usepackage{amssymb}
\usepackage{appendix}
\usepackage{float}
\usepackage{makecell}
\usepackage{amsmath}
\usepackage{txfonts}

\usepackage{hyperref}
\hypersetup{
    unicode, 
    colorlinks=true,
    linkcolor=linkcolor,
    citecolor=linkcolor,
    filecolor=linkcolor,
    urlcolor=linkcolor,
}
\usepackage{color,colortbl}
\usepackage{tensind}
\tensordelimiter{?}
\DeclareGraphicsExtensions{.bmp,.png,.jpg,.pdf}
\usepackage{verbatim}
\usepackage[normalem]{ulem}
\usepackage{orcidlink}
\usepackage{soul}

\DeclareUnicodeCharacter{0306}{}

\newcommand{\orcidauthor}[3]{\author{\href{http://orcid.org/#1}{#2$^{#3}$}}}
  
\graphicspath{ {./figs/} }

\shorttitle{Exo Skryer}
\shortauthors{Elspeth K.H. Lee}

\begin{document}

\title{Exo Skryer: A JAX-accelerated sub-stellar atmospheric retrieval framework.\vspace{-1.5cm}}

\orcidauthor{0000-0002-3052-7116}{Elspeth K.H. Lee}{}
\affiliation{Center for Space and Habitability, University of Bern, Gesellschaftsstrasse 6, CH-3012 Bern, Switzerland}

\begin{abstract}
Contemporary exoplanet and brown dwarf atmospheric research relies heavily on retrieval frameworks to recover thermal and chemical properties and perform model comparison in an observational data-driven approach.
However, the computational effort required for retrieval modelling has rapidly increased, driven by JWST data that covers large spectral intervals at moderate spectral resolutions, and ground-based, high-resolution spectroscopy.
To help tackle the computational burden faced by contemporary retrieval requirements, I present a new sub-stellar atmosphere retrieval modelling framework, \textit{Exo Skryer}, that utilises the JAX library for Python to enable scalable, computationally efficient forward modelling as well as posterior sampling.
I present example retrievals for pre- and current JWST era observations for both transmission and emission spectra, finding consistent results with previous retrieval modelling efforts, apart from a WASP-107b test case. 
In addition, I present a new method to directly retrieve the real and imaginary optical constants ($n$, $k$) of suspected aerosol infrared absorption features.
Due to its computational expediency, \textit{Exo Skryer} will be highly suited for future demanding retrieval efforts that incorporate more spatial dimensionality, complex forward models and high-dimensional parameter sets.
\textit{Exo Skryer} is available as open-source software on GitHub\footnote{\url{https://github.com/ELeeAstro/Exo_Skryer}}.
\end{abstract}


\maketitle


\section{Introduction}
\label{sec:intro}

The information content of exoplanet and brown dwarf atmospheres available for analysis by astronomers has increased substantially since the start of the JWST era, with new insights into the nature of their atmospheres published on a near-weekly basis \citep[e.g.][]{Tsai_2023, Grant_2023, Welbanks_2024}.
In tandem with this slew of new data, retrieval modelling has now become a standard component of first-look analyses of JWST data.
Numerous retrieval codes have been developed in the past couple decades and applied to JWST data, with several listed in \citet{MacDonald_2023b}.
Some examples include \textit{NEMESIS} \citep{Barstow_2014}, \textit{CHIMERA} \citep{Line_2013}, \textit{TauREx} \citep{Waldmann_2015}, \textit{petitRADTRANS} \citep{Molliere_2019}, \textit{Aurora} \citep{Welbanks_2021}, \textit{POSEIDON} \citep{MacDonald_2023}, \textit{ARCiS} \citep{Min_2020}, \textit{BeAR} \citep{Kitzmann_2020}, \textit{Pyrat Bay} \citep{Cubillos_2021}, \textit{PETRA} \citep{Lothringer_2020}, and \textit{Brewster} \citep{Burningham_2017}.

The aim of retrieval in contemporary exoplanet atmospheric research is to statistically infer in a formal manner the atmospheric parameters of the exoplanet utilising a data-driven approach, where the observational data points and errors are used to inform the likelihood of a set of sampled forward model parameters in a statistically consistent approach.
Model comparison can also be made in a systematic way through comparing the ratio of Bayesian evidence \citep[e.g.][]{Kass_1995, Thorngren_2026}.
This is performed by neglecting or adding parts to the forward model from an initial baseline, for example, removing individual gas phase opacity sources one by one \citep[e.g.][]{Felix_2025, Welbanks_2025}.
More advanced model comparison techniques such as `leave one out' (LOO) \citep[e.g.][]{Welbanks_2023} to estimate the sensitivity of the results to individual data points, have also been applied to exoplanet retrieval efforts.

The high performance computing (HPC) infrastructure accessible to typical researchers in astrophysics is increasingly becoming more heterogeneous, including mixtures of central processing units (CPUs) and graphics processing units (GPUs).
This is primarily driven by the rapid development and rollout schedules of machine learning (ML) products in the technology industrial sector, where GPUs are a critical component for scaling the training and inference for these models.
Institute and national HPC infrastructures are increasingly trending with industry in procurement for their HPC inventories.

As both the research interest and utilisation of this infrastructure at research institutes increase, this is making GPUs more commonly available at institutional HPC centres.
This is incentivising scientific codes to migrate from CPUs to GPUs, especially for computationally intensive workloads such as exoplanet atmosphere retrieval modelling \citep[e.g.][]{Molliere_2025, Welbanks_2026}.

In this paper, I present a new atmospheric retrieval framework, \textit{Exo Skryer}, a flexible Python + JAX based code which can operate both the forward model and sampling scheme on CPUs, GPUs or a combination of both. 
I apply \textit{Exo Skryer} to topical 1D transmission and emission spectra scenarios encountered by contemporary sub-stellar retrieval studies.
This paper acts as the core reference for \textit{Exo Skryer}, with further development expected to occur in the next years.
In Section \ref{sec:Exo_Skryer}, I present the \textit{Exo Skryer} retrieval framework, the JAX programming paradigm and nested-sampling options with JAX, the temperature-pressure, chemistry, and cloud structure options within \textit{Exo Skryer}.
In Section \ref{sec:opac}, I detail the various gas phase opacities that \textit{Exo Skryer} can utilise.
In Section \ref{sec:trans}, I detail the radiative-transfer methodology used for the 1D transmission spectra retrievals.
In Section \ref{sec:em}, I detail the radiative-transfer methodology used for the 1D emission spectra retrievals.
In Section \ref{sec:ret}, I present various retrieval examples and results using \textit{Exo Skryer}, spanning pre-JWST and JWST era observational data.
Section \ref{sec:disc} contains the discussion, and Section \ref{sec:conc} the conclusions of this study.

\section{The Exo Skryer Framework}
\label{sec:Exo_Skryer}

\textit{Exo Skryer} is designed with high modularity and flexibility in mind.
The code is driven by a simple YAML-formatted input file, which details the physics modules, opacity sources, retrieval parameters, sampling options and runtime (e.g. number of CPUs/GPUs) environment.
\textit{Exo Skryer} then uses the configuration to build a custom forward model that is then used for the likelihood evaluation inside the sampling framework.
This same YAML file is then used to post-process the posterior samples to produce corner plots, best fit plots and other diagnostics, ensuring the exact same custom forward model is used.
Parameters and any intermediate data are stored as flexible Python dictionaries, which allows data present in the YAML file to be sent to the forward model building routine.
This includes the sampled parameters, constant parameters and other auxiliary data that is required for the specific forward models to be used.
Users can design and integrate their own JAX-compatible physics functions, such as radiative-transfer, chemistry and cloud models, which can be used to build the sampled forward model function.

\subsection{Why JAX?}
\label{sec:why_jax}

\textit{Exo Skryer} is built around JAX\footnote{\url{https://github.com/jax-ml/jax}} \citep{jax2018github}, an open-source numerical computing library for Python that provides a NumPy-like interface with just-in-time (JIT) compilation and accelerator (GPU and TPU) support. 
Retrieval workflows are dominated by repeated forward-model evaluations inside a sampler; even modest improvements in model throughput translate directly into substantial reductions in end-to-end runtime.

JAX offers three advantages that align well with retrieval modelling: (i) compilation of array-based kernels, reducing Python overhead and improving throughput; (ii) straightforward execution on CPUs and GPUs, enabling scalable evaluation over large spectral grids; and (iii) automatic differentiation, which provides a pathway to future gradient-based optimisation and inference, as well as sensitivity analyses.

The main trade-off is that JAX encourages a more functional programming style and requires care with side effects and control flow, which can necessitate refactoring and additional testing. 
In return, it provides a robust foundation for efficient retrieval modelling on increasingly heterogeneous HPC infrastructure.

\subsection{Nested sampling with JAX}

Nested sampling (NS) \citep[e.g.][]{Skilling_2004, Buchner_2023} has become the de facto standard sampling technique for sub-stellar atmosphere retrieval modelling in the past decade \citep[e.g.][]{Welbanks_2021}, mostly outpacing the MCMC based techniques used in earlier retrieval efforts \citep[e.g.][]{Benneke_2012} in day-to-day workloads.
An in-depth description of NS is not in the scope of this study, however, a key benefit of NS is the ability to easily make model comparison and assess detection significance through calculation of the `Bayesian evidence' of different relative parameter sets.

I use three nested sampling frameworks in \textit{Exo Skryer}: \textit{pymultinest} \citep{Buchner_2014}, \textit{dynesty} \citep{Speagle_2020} and \textit{JAXNS} \citep{Albert_2020}.
As a JAX based framework, \textit{JAXNS} can operate on the GPU, enabling both the sampling and forward model stages to be performed on the GPU.
\textit{pymultinest} and \textit{dynesty} cannot utilise the GPU at the sampling stage, but benefit instead from the JAX forward model, which is able to be performed on the GPU, or multi-threaded CPU.
This enables a hybrid approach similar to the \textit{BeAR} retrieval model \citep{Kitzmann_2020}, where the sampling stage is performed on the CPU, with each likelihood evaluation performed on the GPU.
In addition, the JAX infrastructure also accelerates CPU only workloads, with \textit{Exo Skryer} able to run without the need for GPUs.
This allows practitioners without GPU access to also run performant retrievals.
The \textit{ExoJAX} \citep{Kawahara_2022, Kawahara_2025} model is another example of utilising JAX to perform retrievals on sub-stellar atmosphere data.
\textit{ExoJAX} is highly optimised for performing MCMC calculations with highly accurate medium-to-high-resolution spectral forward modelling.
\textit{ExoJAX} is also fully differentiable, enabling highly efficient parameter exploration with advanced MCMC methods such as No-U-Turn Sampler (NUTS).

\subsubsection{Speed testing samplers}
\label{sec:speed}

\begin{table}[]
    \centering
    \caption{Retrieval completion time (minutes) for the different nested-sampling options in \textit{Exo Skryer} for the HD 209458b pre-JWST retrieval example.}
    \begin{tabular}{c|c|c|c} \hline \hline
        Scheme & ck ($R$=1000) & OS ($R$=10,000) & OS ($R$=20,000) \\ \hline
        \textit{pymultinest} & 23 & 18 &  28\\
        \textit{dynesty} & 63 & 40 &  66 \\
        \textit{JAXNS} & 324 & 252& 421 \\ \hline \hline
    \end{tabular}
    \label{tab:speed}
\end{table}

To compare the computational efficiency of a nominal setup of \textit{Exo Skryer}, I performed the HD 209458b pre-JWST retrieval setup (11 parameters) from Section \ref{sec:HD209} with \textit{pymultinest}, \textit{dynesty} and \textit{JAXNS} utilising the same JAX forward model setup, each using correlated-k (ck) at a spectral resolution of $R$ = 1,000 and opacity sampling (OS) at spectral resolutions of $R$ = 10,000 and $R$ = 20,000.
1000 live points and a stopping criterion of $\Delta\ln Z$ = 0.1 were used in all cases.
\textit{pymultinest} and \textit{dynesty} use the hybrid CPU-GPU setup, with one CPU and one GPU for the computation, while the \textit{JAXNS} test performs all sampling and forward model calculations on the GPU.
Table \ref{tab:speed} presents the results of this test, where \textit{pymultinest} is the most efficient and \textit{JAXNS} the least efficient.

For the \textit{JAXNS} test, a plausible explanation is that the consumer-grade GPU used in this study (Nvidia RTX 4090) may be too saturated performing the sampling as well as the forward model, which significantly degrades performance.
Owing to this speed testing and recommendations from previous comparison studies \citep[e.g.][]{Speagle_2020}, I use \textit{dynesty} for all retrieval models in this study.
Specifically, I use the hybrid CPU-GPU setup, utilising \textit{dynesty} with one CPU and the JAX accelerated forward model with one GPU.

\subsection{Likelihood}

I use a standard independent Gaussian errors formulation for the log-likelihood function \citep[e.g.][]{Kitzmann_2020}
\begin{equation}
\ln \mathcal{L}
=
\sum_{i=1}^{N}
\left[
-\log \sigma_{{\rm eff},i}
-\frac{1}{2}\log(2\pi)
-\frac{1}{2}\left(\frac{r_i}{\sigma_{{\rm eff},i}}\right)^2
\right],
\end{equation}
where $r_{i}$ is the residual between the observational data and forward model for the $i$th observational point
\begin{equation}
    r_{i} = y_{i}^{\rm obs} - y_{i}^{\rm model},
\end{equation}
and $\sigma_{\rm eff}$ the effective 1$\sigma$ errors in the observational point.
Any invalid outputs from the forward model are discarded with a log-likelihood of $\log\mathcal{L}$ = -10$^{300}$.

For brown dwarf retrievals, I include an error inflation parameter, $c$, in the form
\begin{equation}
\label{eq:c_error}
    \sigma_{\rm eff}^{2} = \sigma^{2} + 10^{2c},
\end{equation}
with the prior bounds 
\begin{equation}
    \log_{10}\left[0.1 \times {\rm min}(\sigma)\right] < c < \log_{10}\left[10 \times {\rm max}(\sigma)\right],
\end{equation}
where $\sigma$ is the 1$\sigma$ uncertainty in the observational data.
For retrievals that don't require inflation, I force $c$ = -99, effectively setting zero error inflation.

This scheme is similar to the commonly used error inflation following \citet{Tremaine_2002, Hogg_2010} and \citet{Foreman-Mackey_2013}, for example, in \citet{Line_2015} and \citet{Burningham_2017}, which used the equation
\begin{equation}
    \sigma_{\rm eff}^{2} = \sigma^{2} + 10^{b},
\end{equation}
with the prior bounds
\begin{equation}
    \log_{10}\left[0.01 \times {\rm min}(\sigma^{2})\right] < b < \log_{10}\left[100 \times {\rm max}(\sigma^{2})\right].
\end{equation}
However, I find utilising Eq. \ref{eq:c_error} enables a simpler and more intuitive setting of the error inflation priors, using factors of the $\sigma$ directly which scale more cleanly to the published error in the observational data, rather than factors of $\sigma^{2}$.
This allows a more direct comparison between the posterior error inflation in the retrieval and the 1$\sigma$ observational errors.

\subsection{$T$-$p$ profile parameterisation}

Every retrieval framework must offer a simple parameterisation of the $T$-$p$ profile that is quick to evaluate, and provides the first step in performing the forward model.
Below I present some options available in \textit{Exo Skryer} used in the current study.

\subsubsection{Isothermal}

The simplest $T$-$p$ profile, and commonly used in transmission spectra retrieval efforts, is assuming the entire atmosphere is isothermal throughout the vertical extent.
This leads to only $T_{\rm iso}$ [K] as the retrieved variable for this structure
\begin{equation}
    T(p) = T_{\rm iso}.
\end{equation}
I use this profile in the WASP-17b transmission spectra retrieval in Section \ref{sec:W17b}.

\subsubsection{Barstow 2020}

A simple, but more physical option was used in \citet{Barstow_2020}, where a `stratospheric' isothermal temperature, $T_{\rm strat}$ [K], is chosen with an adiabatic profile between 0.1-1 bar, and then an isothermal deep pressure region
\begin{equation}
       T(p) = \begin{cases}
      T_{\rm strat}, &  p \le 0.1, \\
      T_{\rm strat} \left(\frac{p}{0.1}\right)^{\kappa}, & 0.1 < p < 1, \\
      T_{\rm strat} \left(\frac{1}{0.1}\right)^{\kappa},  & p \ge 1,
   \end{cases}
\end{equation}
where $\kappa$ $\approx$ 2/7 is the adiabatic constant.
Again, only a single parameter, $T_{\rm strat}$ [K], is required in the retrieval model.
I use this profile in the HD 209458b pre-JWST transmission spectra retrieval setup in Section \ref{sec:HD209}.

\subsubsection{Modified Milne}
\label{sec:Milne}

\begin{figure}
    \centering
    \includegraphics[width=1.0\linewidth]{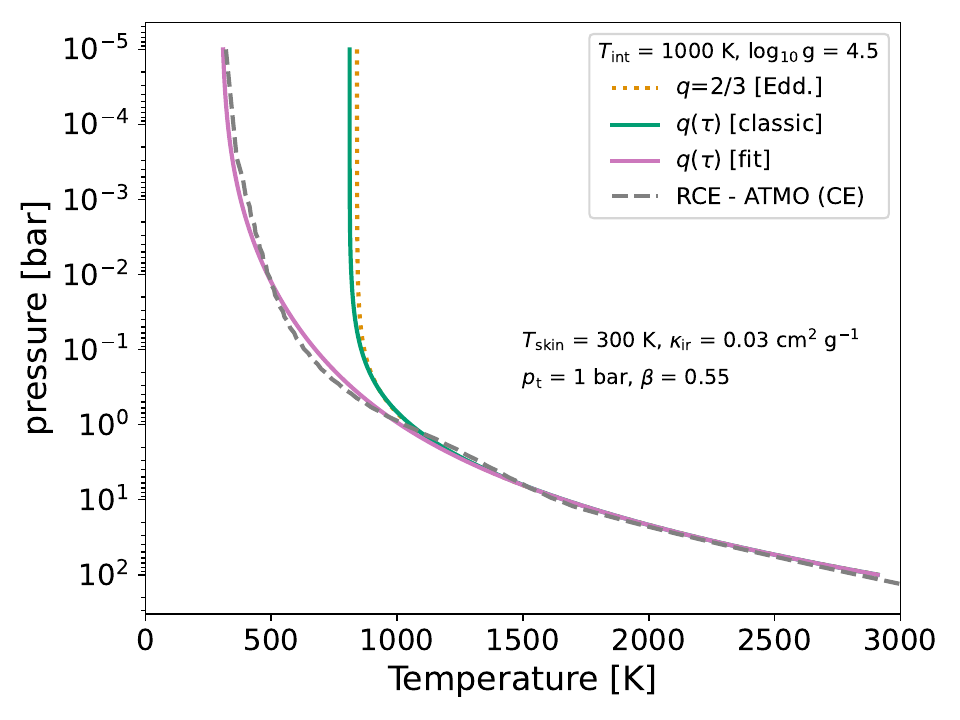}
    \caption{Example of the modified Milne $T$-$p$ profile scheme for brown dwarf retrievals in Sect. \ref{sec:Milne} (purple, solid), compared to the original Milne profile (orange, dotted), classic Hopf solution (green, solid) and a chemical equilibrium ATMO profile (grey, dashed) from \citet{Phillips_2020}.
    The bulk parameters are $T_{\rm int}$ = 1000 K, $\log_{10}$ g = 4.5, with the modified Milne profile parameters printed inside the figure.}
    \label{fig:Milne}
\end{figure}

For isolated exoplanet and brown dwarf atmospheres, due to the lack of irradiation, the $T$-$p$ structure is expected to be more monotonic and driven primarily by the internal temperature.
An option is to use the classic grey Milne analytical radiative-equilibrium (RE) solution
\begin{equation}
    T^{4}(\tau) = \frac{3}{4}T_{\rm int}^{4}\left[\tau + q(\tau)\right],
\end{equation}
where $q(\tau)$ is the Hopf function, which is $q(\tau)$ = 2/3 when taking the Eddington approximation for isotropic, diffuse emission.
The optical depth, $\tau$, of the grey atmosphere is given by 
\begin{equation}
    \tau(p) = \frac{\kappa_{\rm ir}}{g}p,
\end{equation}
where $\kappa_{\rm ir}$ [cm$^{2}$ g$^{-1}$] is the constant infrared band opacity of the atmosphere, $p$ [dyne cm$^{-2}$] the atmospheric pressure and $g$ [cm s$^{-2}$] the gravitational acceleration. 

For retrieval purposes, I propose a parameterised Hopf function that takes the form of a stretched exponential function
\begin{equation}
\label{eq:q}
    q(\tau) = q_{\infty} + (q_{0} - q_{\infty})\exp\left[-\left(\frac{\tau}{\tau_{\rm t}}\right)^{\beta}\right],
\end{equation}
which can be transposed to pressure coordinates when $\kappa_{\rm ir}$ is constant
\begin{equation}
    q(p) = q_{\infty} + (q_{0} - q_{\infty})\exp\left[-\left(\frac{p}{p_{\rm t}}\right)^{\beta}\right],
\end{equation}
where $q_{0}$ is the value of $q$ as $\tau$ $\rightarrow$ 0, $q_{\infty}$ the value of $q$ as $\tau$ $\rightarrow$ $\infty$, $p_{\rm t}$ [dyne cm$^{-2}$] the transition pressure and $\beta$ a stretching parameter.
Physically, $\tau_{\rm t}$ or $p_{\rm t}$ marks the approximate boundary between the deep and upper $q$ dominance, and $\beta$ controls the sharpness of the transition, with $\beta$ $>$ 1 being stronger and $\beta$ $<$ 1 being weaker than an exponential transition respectively.
The parameter $q_{0}$ is related to the skin temperature of the atmosphere through
\begin{equation}
  q_{0} = \frac{4}{3}\left(\frac{T_{\rm skin}}{T_{\rm int}}\right)^{4},
\end{equation}
a suitable retrieval parameter is then the skin to internal temperature ratio, $T_{\rm ratio}$ (0,1), given as
\begin{equation}
    T_{\rm ratio} = \frac{T_{\rm skin}}{T_{\rm int}}.
\end{equation}
Here, a possibility is to use the classic grey atmosphere solutions at $\tau$ = $\infty$ ($q_{\infty}$ $\approx$ 0.71) and $\tau$ = 0 ($q_{0}$ $\approx$ 1$/\sqrt{3}$) \citep[e.g.][]{Mihalas_1978}.
However, this classic solution was found in \citet{Kitzmann_2020} not to be flexible enough for general brown dwarf retrieval modelling.
I therefore relax constraints on $q_{0}$ to allow greater flexibility in the $T$-$p$ structure, and the ability to recover a $T$-$p$ profile more akin to non-grey RCE simulation results.

In Figure \ref{fig:Milne}, I show an example fit to an ATMO \citep{Phillips_2020} RCE profile using this scheme, as well as the Eddington and classic Hopf function solution $T$-$p$ profiles.
The modified Milne scheme is able to well capture the $T$-$p$ structure of the RCE model output, while the classic solutions turn off into an isothermal upper atmosphere too quickly, and fail to capture the decreasing temperature gradient from the RCE model.
Overall, the required retrieval parameters for this scheme are $T_{\rm int}$ [K], $\kappa_{\rm ir}$ [cm$^{2}$ g$^{-1}$], $T_{\rm ratio}$, $p_{\rm t}$ [dyne cm$^{-2}$] and $\beta$ for five total.
I use this $T$-$p$ profile for the Gliese 229B retrieval in Section \ref{sec:G229B}.

\subsubsection{Modified Guillot}
\label{sec:Guillot}

\begin{figure}
    \centering
    \includegraphics[width=1.0\linewidth]{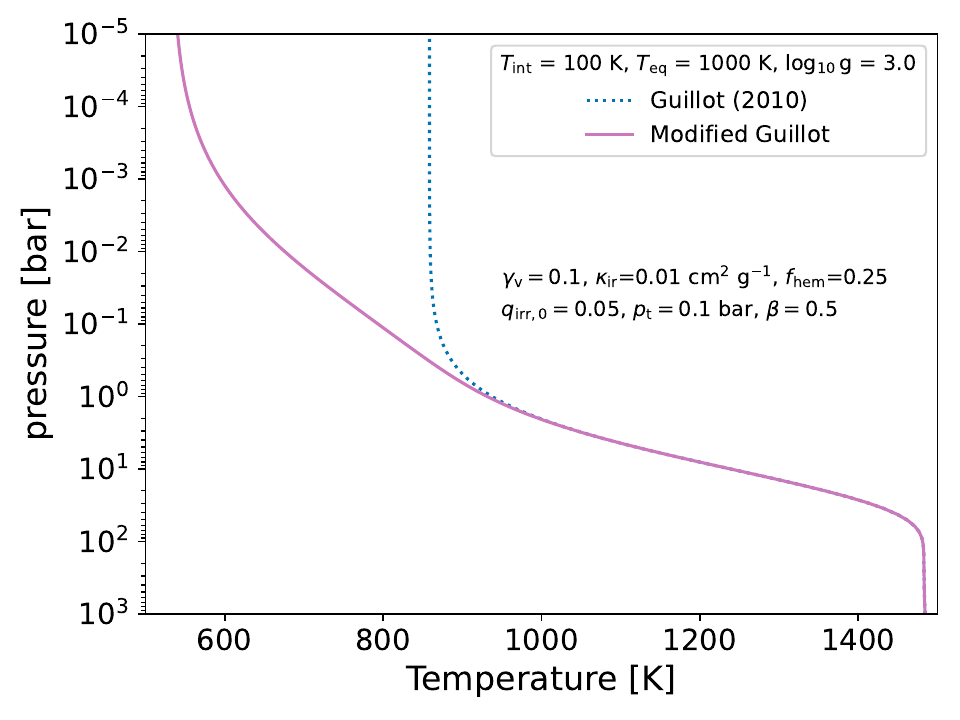}
    \caption{Example of the modified Guillot $T$-$p$ profile scheme for hot Jupiter retrievals in Sect. \ref{sec:Guillot} (purple, solid), compared to the original \citet{Guillot_2010} profile (blue, dotted).
    The bulk parameters are $T_{\rm int}$ = 100 K, $T_{\rm eq}$ = 1000 K, $\log_{10}$ g = 3.0, with the modified Guillot profile parameters printed inside the figure.}
    \label{fig:Guillot}
\end{figure}

A commonly utilised $T$-$p$ profile parameterisation for hot Jupiters are the several radiative-equilibrium (RE), semi-grey analytical solutions derived in \citet{Guillot_2010}.
The main $T$-$p$ solution in \citet{Guillot_2010} is given by
\begin{equation}
\begin{split}
    T^{4}(\tau) &= \frac{3T_{\rm int}^{4}}{4}\left[\frac{2}{3} + \tau\right] \\ 
     & + \frac{3T_{\rm irr}^{4}}{4}\mu_{\star} \left[\frac{2}{3} + \frac{\mu_{\star}}{\gamma_{\rm v}} + \left(\frac{\gamma_{\rm v}}{3\mu_{\star}} - \frac{\mu_{\star}}{\gamma_{\rm v}}\right)e^{-\gamma_{\rm v}\tau/\mu_{\star}}\right], 
\end{split}
\end{equation}
where $T_{\rm int}$ [K] is the internal temperature, $T_{\rm irr}$ [K] the irradiation temperature, $\gamma_{\rm v}$ the ratio between the visible and infrared grey opacity and $\mu_{\star}$ the stellar zenith angle.

However, the \citet{Guillot_2010} profile tends toward an isothermal limit at small optical depth under the semi-grey, constant-opacity assumptions.
This isothermal behaviour has been shown, when used in retrieval and $T$-$p$ recovery contexts, to produce potentially unreliable results \citep[e.g.][]{Blecic_2017, Barstow_2020b, Dobbs-Dixon_2022}.
Following a similar idea to the modified Milne profile in Section \ref{sec:Milne}, I propose modifying the Guillot profile utilising a flexible Hopf-like function, where, for a global mean $T$-$p$ profile, I assume $\mu_{\star}$ = 1/$\sqrt{3}$ and
\begin{equation}
\begin{split}
    T^{4}(\tau) &= \frac{3T_{\rm int}^{4}}{4}\left[q_{\rm int}(\tau) + \tau\right] \\ 
     & + \frac{3T_{\rm eq}^{4}}{4}4f_{\rm hem} \left[q_{\rm irr}(\tau) + \frac{1}{\gamma_{\rm v}\sqrt{3}} + \left(\frac{\gamma_{\rm v}}{\sqrt{3}} - \frac{1}{\gamma_{\rm v}\sqrt{3}}\right)e^{-\gamma_{\rm v}\tau\sqrt{3}}\right], 
\end{split}    
\end{equation}
where $q_{\rm int}$ is the parameterised Hopf function for the internal radiation, and $q_{\rm irr}$ the parameterised Hopf function for the incident radiation, where both use the stretched exponential form in Eq. \eqref{eq:q}.
$f_{\rm hem}$ [0, 1] is the hemispheric energy redistribution factor \citep[e.g.][]{Guillot_2010}, which typically takes values of 1 (sub-stellar point), 0.5 (dayside redistribution), and 0.25 (4$\pi$ hemisphere redistribution).
In this study, I assume a constant $f_{\rm hem}$ = 0.25.

Since for hot Jupiters $T_{\rm eq}^{4}$ $\gg$ $T_{\rm int}^{4}$, the influence of the internal flux Hopf function is small and $q_{\rm int}$ can be returned to the Eddington approximation, $q_{\rm int}$ = 2/3, lowering the number of overall retrieved parameters.
The irradiated flux Hopf function has the most impact on the modified Guillot $T$-$p$ profile.
For recovering suitable retrieval parameters, I forgo relating $q_{\rm irr, 0}$ to the skin temperature as in the modified Milne profile, instead directly retrieving 0 $<$ $q_{\rm irr, 0}$ $\le$ 2/3 and setting $q_{\rm irr, \infty}$ = 2/3.
For ultra hot Jupiters, where more complex temperature inversion behaviour is desirable, retrieving 0 $<$ $q_{\rm irr, 0}$ $<$ 1 can offer even more flexibility.
In Figure \ref{fig:Guillot}, I show an example of this modified $T$-$p$ parameterisation, which shows the added upper atmosphere flexibility of this scheme compared to the original \citet{Guillot_2010} formulation.
Overall, the required retrieval parameters for this scheme are $T_{\rm int}$ [K], $T_{\rm eq}$ [K], $\kappa_{\rm ir}$ [cm$^{2}$ g$^{-1}$], $\gamma_{\rm v}$, $q_{\rm irr, 0}$, $p_{\rm t}$ [dyne cm$^{-2}$] and $\beta$ for seven total if $f_{\rm hem}$ is kept fixed.

I note that these additional parameterisations break the RE solution presented in \citet{Guillot_2010}, similar to the modified Milne profile, and so retrieved profiles should be used with caution when physically interpreting parameters derived from retrievals. 
However, this parameterisation offers a more flexible, retrieval-friendly generalisation of the Guillot profile that can emulate non-grey upper-atmosphere behaviour while retaining the deep-atmosphere RE solution.
I use this profile for the WASP-107b transmission spectra in Section \ref{sec:W107b} and HD 189733b dayside emission spectra retrieval in Section \ref{sec:HD189b}.

\subsection{Non-grey aerosol opacity}

Including the effects of haze and cloud (aerosol) opacity in a parameterised manner that is also physically interpretable is a key goal of contemporary retrieval modelling efforts \citep[e.g.][]{Wakeford_2017, Burningham_2017, Molliere_2020, Burningham_2021, Vos_2023, Dyrek_2024, Nasedkin_2025, Molliere_2025}.
In particular, with the wide wavelength coverage available from JWST observations, understanding the effects of cloud opacity across the whole spectral range is now an important consideration.
Below, I detail the cloud models used in this study, focusing on non-grey formulations.

\subsubsection{Vertical cloud profiles}

In \textit{Exo Skryer}, the basic unit for the vertical cloud structures is the mass mixing ratio, $q_{\rm c}$ [g g$^{-1}$], of the condensate
\begin{equation}
    q_{\rm c} = \frac{\rho_{\rm c}}{\rho_{\rm a}},
\end{equation}
where $\rho_{\rm c}$ [g cm$^{-3}$] is the mass density of the condensate and $\rho_{\rm a}$ [g cm$^{-3}$] the mass density of the background atmosphere.
I use three schemes in this study: a vertically uniform slab profile across the entire vertical extent
\begin{equation}
 q_{\rm c}(p) = q_{\rm c, slab},
\end{equation}
with only one parameter: $q_{\rm c, slab}$ [g g$^{-1}$],
a uniform slab defined between a top pressure and a given pressure range \citep[e.g.][]{Mullens_2024}
\begin{equation}
 q_{\rm c}(p) =
\begin{cases}
            0,        & p < p_{\rm c, top}, \\
    q_{\rm c, slab},  &  p_{\rm c, top} \le p \le p_{\rm c, top} + \Delta p, \\
            0,        & p > p_{\rm c, top} + \Delta p, \\
\end{cases}
\end{equation}
which has three parameters: $q_{\rm c, slab}$ [g g$^{-1}$], $p_{\rm c, top}$ [dyne cm$^{-2}$] and $\Delta p$ [dyne cm$^{-2}$],
and a vertical mass mixing ratio profile with a base value, $q_{\rm c, base}$ [g g$^{-1}$], given at a cloud base pressure, $p_{\rm c, base}$ [dyne cm$^{-2}$], and a power law relation, $\alpha$, that reduces the mass mixing ratio with fractions of the pressure scale height
\begin{equation}
    q_{\rm c}(p) = q_{\rm c, base} \left(\frac{p}{p_{\rm c, base}}\right)^{\alpha},
\end{equation}
where $q_{\rm c}(p)$ = 0 for $p$ $>$ $p_{\rm c, base}$.
This mimics the cloud structures typically seen in forward modelling efforts \citep[e.g.][]{Ackerman_2001, Ohno_2017, Batalha_2026} and has a total of three parameters: $q_{\rm c, base}$ [g g$^{-1}$], $p_{\rm c, base}$ [dyne cm$^{-2}$] and $\alpha$.

\subsubsection{Fisher \& Heng (2018)}
\label{sec:F18}

\citet{Fisher_2018} propose a non-grey parameterisation of the cloud extinction opacity, $\kappa_{\rm c, ext}$ [cm$^{2}$ g$^{-1}$], as
\begin{equation}
    \kappa_{\rm c, ext} = \frac{\kappa_{0}}{Q_{0}x^{-a} + x^{0.2}},
\end{equation}
where $\kappa_{0}$ [cm$^{2}$ g$^{-1}$] is an effective baseline grey cloud opacity, $Q_{0}$ related to the peak of the extinction efficiency curve, $a$ a spectral slope index, and $x$ = 2$\pi$$r$/$\lambda$ the size parameter.
The retrieved parameters for this scheme are therefore $\kappa_{0}$ [cm$^{2}$ g$^{-1}$], $Q_{0}$, $r$ [cm], and $a$.
This formulation was compared to other cloud opacity schemes in \citet{Barstow_2020}.

I propose a slight variation on the \citet{Fisher_2018} formulation to recover a more physically interpretable unit than the baseline cloud opacity.
First, I use the equation for the cloud number density, $N_{0}$ [cm$^{-3}$], for a monodisperse size distribution, assuming spherical particles
\begin{equation}
    N_{0} = \frac{3\rho_{\rm a}q_{\rm c}}{4\pi\rho_{\rm d}r_{\rm eff}^{3}},
\end{equation}
where $q_{\rm c}$ [g g$^{-1}$] is the mass mixing ratio of the cloud particles, $\rho_{\rm a}$ [g cm$^{-3}$] the density of the background atmosphere, $r_{\rm eff}$ [cm] the effective particle size, and $\rho_{\rm d}$ [g cm$^{-3}$] the bulk density of the cloud material.
The opacity of the cloud particles is then approximated by
\begin{equation}
    \kappa_{\rm c, ext} \approx \frac{N_{0}Q_{\rm ext}(r_{\rm eff})\pi r_{\rm eff}^{2}}{\rho_{\rm a}} = \frac{3q_{\rm c}Q_{\rm ext}(r_{\rm eff})}{4\rho_{\rm d}r_{\rm eff}},
\end{equation}
where, for convenience and computational expediency, the same fitting form to \citet{Kitzmann_2018} for $Q_{\rm ext}$ can be assumed 
\begin{equation}
\label{eq:Q_ext_KH}
    Q_{\rm ext} = \frac{Q_{1}}{Q_{0}x^{-a} + x^{0.2}}.
\end{equation}
To reduce the number of retrieved parameters, I subsume the variables $Q_{1}$ and $\rho_{\rm d}$ into $q_{\rm c}$ to create an effective mass mixing ratio, $q_{\rm c}^{\rm eff}$
\begin{equation}
    q_{\rm c}^{\rm eff} = \frac{Q_{1}}{\rho_{\rm d}}q_{\rm c}.
\end{equation}
For typical silicate materials $Q_{1}$ $\sim$ $\rho_{\rm d}$ \citep{Kitzmann_2018}, and so I expect the subsumed scaling to only be a small factor in the range $\sim$ 1-10.
Alternatively, representative constant values for $Q_{1}$ and $\rho_{d}$ for specific materials can be chosen \citep[e.g. see][]{Kitzmann_2018}.
This leaves the retrieved parameters to be: $r_{\rm eff}$, $a$, $Q_{0}$ and $q_{\rm c}^{\rm eff}$,  the same number as the original \citet{Fisher_2018} scheme.
This scheme can be easily extended to polydisperse size distributions, such as the log-normal distribution.
However, this is not used in the current study.

To reproduce a wavelength-independent grey cloud, the geometric optics limit ($x$ $\gg$ 1), $Q_{\rm ext}$ = 2, can be assumed, which reduces the number of parameters to only $r_{\rm eff}$ and $q_{\rm c}$.
This allows a simple way to recover representative radii and mass mixing ratios without assuming any wavelength dependence on the cloud opacity, though practitioners wanting to further reduce the number of sampled parameters may wish to retain a single grey cloud opacity formulation.

For retrieving cloudy versus cloud-free atmosphere fractions, I use the simple linear combination of spectra with and without the effects of cloud opacity, for transmission and emission spectra \citep{Marley_2010, Line_2016}
\begin{equation}
    \Phi = \Phi_{\rm cld}f_{\rm cld} + (1 - f_{\rm cld})\Phi_{\rm cf},
\end{equation}
where $f_{\rm cld}$ [0, 1] is the fraction of the spectra that contains cloud opacity, $\Phi_{\rm cld}$ the spectra including cloud opacity and $\Phi_{\rm cf}$ the cloud free spectra without the effects of cloud opacity.

\subsubsection{Adding mid-infrared absorption features}
\label{sec:feat}

\begin{figure}
    \centering
    \includegraphics[width=1.0\linewidth]{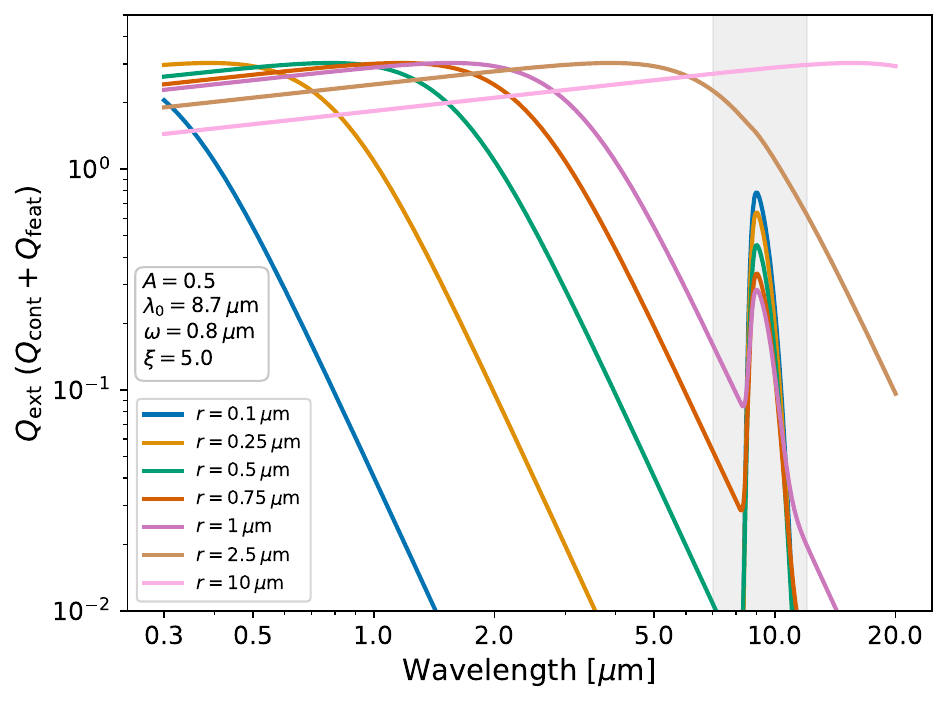}
    \caption{Extinction efficiency with wavelength for SiO$_{2}$ applying the \citet{Kitzmann_2018} scheme for the continuum (Eq. \ref{eq:Q_ext_KH}) combined with the parameterised mid-infrared absorption feature extinction efficiency (Eq. \ref{eq:Q_ext_feat}).
    Various particle radii are shown, with the assumed skewed Gaussian parameters shown as text in the plot.
    The background grey highlighted region shows the typical wavelength range where silicate absorption features are located.}
    \label{fig:Q_ext_feat}
\end{figure}

The \citet{Fisher_2018} and \citet{Kitzmann_2018} scheme is only able to capture the wavelength dependent continuum opacity of the clouds, and does not approximate the mid-infrared absorption features.
I therefore propose a, retrieval friendly, addition to the \citet{Fisher_2018} scheme based on the skewed Gaussian feature parameterisation found in \citet{Welbanks_2024} which was used to fit the cloud absorption feature in WASP-107b.
The total extinction efficiency is given by
\begin{equation}
    Q_\mathrm{ext}(\lambda) = \min\!\Bigl[
        Q_\mathrm{cont}(\lambda) + W(x)\,Q_\mathrm{feat}(\lambda),\;
        Q_\mathrm{max}
    \Bigr],
\end{equation}
where $Q_\mathrm{max} = 2$ guards against extreme unphysical values.
The analytical continuum extinction efficiency, $Q_\mathrm{cont}$, is given by the \citet{Kitzmann_2018} parameterisation (Eq. \ref{eq:Q_ext_KH}).
The parameterised absorption feature extinction efficiency, $Q_\mathrm{feat}$, is a skew-normal profile in a given wavelength range following \citet{Welbanks_2024},
\begin{equation}
\label{eq:Q_ext_feat}
    Q_\mathrm{feat}(\lambda) =
        2A \exp\!\left(-\tfrac{1}{2}z^2\right)
        \Phi(\xi z),
    \qquad
    z = \frac{\lambda - \lambda_0}{\omega},
\end{equation}
where $\Phi(z) = \tfrac{1}{2}\bigl[1 + \mathrm{erf}(z/\sqrt{2})\bigr]$
is the standard normal cumulative distribution function (CDF), $A$ is the peak amplitude, $\lambda_0$ the
central wavelength, $\omega$ the width, and $\xi$ the skewness parameter.
In the symmetric limit, $\xi \to 0$, the profile reduces to a Gaussian with
amplitude $A$.

I use a size-parameter dependent window function, $W(x)$, to suppress the spectral feature extinction as particles enter the geometric-optics regime 
\begin{equation}
    W(x) = \exp\!\left(-\frac{x}{x_0}\right),
\end{equation}
where $x_{0}$ is a transition size parameter, which is chosen to be 0.5.
In the Rayleigh limit, $x \ll x_0$, the window approaches unity and the
feature is fully present; as $x \to \infty$ the feature is suppressed,
reflecting the washing-out of the absorption feature in the geometric-optics
regime \citep[e.g.][]{Wakeford_2015}.
I note that in this scheme, I assume that $Q_{\rm feat}$ only captures the absorption efficiency, neglecting any scattering contribution to the mid-infrared feature.
In Figure \ref{fig:Q_ext_feat}, I show a visualisation of the resulting $Q_{\rm ext}$ for a set of radii and skewed Gaussian parameters. 

\subsubsection{Direct $n$,$k$ retrieval}
\label{sec:direct_nk}

I propose a new aerosol opacity parameterisation with the ability to quickly estimate the real ($n$) and imaginary ($k$) optical constants of the cloud particles directly.
This allows the retrieval calculation to remain agnostic to the composition of the cloud particles, and compare the wavelength dependent structure of the retrieved optical constants to available databases.
This scheme is an alternative to assuming a specific condensate material and sourcing $n$ and $k$ coefficients published in the literature for that species, and calculating cloud opacities using these optical constants during retrieval  \citep[e.g.][]{Molliere_2025}.

In a retrieval context, I propose combining known analytic solutions to Mie theory to determine $Q_{\rm ext}$ wavelength dependently.
I split the calculation into two size parameter regime approximations, Rayleigh for $x$ $\leq$ 1 and Modified Anomalous Diffraction Theory (MADT) for $x$ $\geq$ 3.
In the Rayleigh regime, the extinction efficiency is the sum of the scattering ($Q_{\rm sca}$) and absorption ($Q_{\rm abs}$) efficiencies given by \citep{Bohren_1983}
\begin{equation}
    Q_{\rm sca} = \frac{8}{3}x^{4}\,\Re\!\left[\left(\frac{m^{2}-1}{m^{2}+2}\right)^{2}\right],
\end{equation}
and
\begin{equation}
    Q_{\rm abs} = 4x\,\Im\!\Bigg[
\left(\frac{m^{2}-1}{m^{2}+2}\right)
\Bigg(
1+\frac{x^{2}}{15}\left(\frac{m^{2}-1}{m^{2}+2}\right)\frac{m^{4}+27m^{2}+38}{2m^{2}+3}\Bigg)\Bigg],
\end{equation}
where $m$ = $n$ + $ik$.
For MADT, I follow the scheme described in \citet{Moosmuller_2018} with
\begin{equation}
    Q_{\rm ext, MADT} = (1 + 0.5C_{2}) Q_{\rm ext, ADT} + Q_{\rm edge},
\end{equation}
and
\begin{equation}
    Q_{\rm abs, MADT} = (1 + C_{1} + C_{2})Q_{\rm abs, ADT},
\end{equation}
where $C_{1}$, $C_{2}$ and $Q_{\rm edge}$ are given in \citet{Moosmuller_2018} and $Q_{\rm ADT}$ are the values derived from Anomalous Diffraction Theory (ADT).

A problem with this approach is the inaccuracy of both Rayleigh and MADT in the intermediate size parameter regime 1 $\lesssim$ $x$ $\lesssim$10, though MADT is a significant improvement over ADT in this regime \citep{Moosmuller_2018}.
To tackle this, I use a scaled smootherstep function to interpolate between the Rayleigh and MADT results between the range 1$<$ $x$ $<$ 3, where the [0,1] smootherstep function is given by
\begin{equation}
    S_{2}(x) = 6x^{5} - 15x^{4} + 10x^{3},
\end{equation}
which is re-scaled between 1 and 3.

Retrieving the individual $n$ and $k$ for each band is computationally infeasible, increasing the number of parameters to 100-1000s and likely to be highly degenerate and uninformative.
I therefore use a parameterised, functional approach to reduce the number of parameters in the retrieval model, recovering the broad trends in $n$ and $k$.
For the $n$ and $k$ constants, I retrieve 12 spline nodes across the wavelength range of interest.
These spline nodes are then used with a shape-preserving Piecewise Cubic Hermite Interpolation Polynomial (PCHIP) scheme to interpolate the $n$ and $k$ values across the wavelength range.
Careful placement in wavelength is required for these nodes, with one required at both extremities of the wavelength range of interest.
Placement can be more targeted through performing a cloud free retrieval first, then considering wavelength regions where suspected cloud absorption features may be present.
For example, in the WASP-17b test case (Section \ref{sec:W17b}), I aim to try to recover the MIRI silicate absorption feature, and so concentrate the nodes around 7-12 $\mu$m.
I avoid placing nodes at the exact wavelengths of the observed data to discourage overfitting, instead placing them at regular intervals across the wavelength range and allowing the smooth interpolation scheme to provide the structure of the ($n$, $k$) coefficients.

I note that in this scheme, nodes are required to be spaced at larger interval spacings than gas phase features, to allow a smooth, continuous interpolation across wavelength, to avoid fitting gas phase features with a sharp cloud opacity feature.
Overall, despite being an analytical approach, I find this method adds significant extra computational work ($\sim$ $\times$10-100) during retrieval, slowing down convergence compared to the simpler fitting function from \citet{Kitzmann_2018} and \citet{Fisher_2018}.

\section{Gas phase opacities}
\label{sec:opac}

\textit{Exo Skryer} has the ability to either use direct, `opacity sampled' (OS) cross-section data or assume correlated-k (ck) with k-tables.
I have included the functionality to read in the data formats used by the \textit{TauREx 3} \citep{Al-Refaie_2021} and \textit{petitRADTRANS} \citep{Molliere_2019} retrieval models available on the ExoMol\footnote{\url{https://www.exomol.com/}} website \citep{Chubb_2021}.
In addition, I provide Python scripts to create custom OS or ck tables from the DACE\footnote{\url{https://dace.unige.ch/opacityDatabase/}} database.
To mix individual species k-tables, I use the random overlap with resampling and rebinning (RORR) method \citep[e.g.][]{Amundsen_2017} or, for transmission spectra, assuming random overlap through multiplication of the transmission functions of each species \citep[e.g.][]{Amundsen_2017}.

In this study, I mostly use a set of custom OS tables created using the \textit{HELIOS-k} opacity calculator \citep{Grimm_2015} at a constant spectral resolution of $R$ = 20,000 between 0.3 $\mu$m and 30 $\mu$m,  with a temperature range of 50-6100 K and a pressure range of 10$^{-8}$-1000 bar.
I include opacity data of H$_{2}$O \citep{Polyansky_2018}, CO$_{2}$ \citep{Yurchenko_2020}, CO \citep{Li_2015}, CH$_{4}$ \citep{Yurchenko_2024}, NH$_{3}$ \citep{Coles_2019}, H$_{2}$S \citep{Azzam_2016}, SO$_{2}$ \citep{Underwood_2016}, Na \citep{Kurucz_1995} and K \citep{Kurucz_1995} where the citation associated with each molecule is the line-list reference used in \textit{HELIOS-k} to produce the high-resolution ($\Delta$$\nu$ = 0.01 cm$^{-1}$) original cross-section tabulated data. 

\subsection{Free chemistry VMR retrieval}

\textit{Exo Skryer} is able to perform `free' retrievals assuming a constant vertical volume mixing ratio (VMR) of gas phase opacity sources, as well as implicitly assuming an H$_{2}$ and He dominated atmospheric composition.
This is in contrast to `chemical equilibrium' (CE) retrievals \citep[e.g.][]{Kitzmann_2020} which assume chemical equilibrium throughout the atmosphere using calculated abundances for species derived from codes such as \textit{Fastchem} \citep{Stock_2022}.

I follow the scheme of \citet{Welbanks_2021} where the remainder of the gas VMR is composed of H$_{2}$ and He background gas, ensuring that $\sum_{i}$$X_{i}$ = 1, given by
\begin{equation}
X_{\mathrm{H_2}}
=
\frac{1 - \sum_{i,\,i\neq \mathrm{He},\mathrm{H_2}}^{n} X_i}
     {1 + \frac{X_{\mathrm{He}}}{X_{\mathrm{H_2}}}},
\qquad
X_{\mathrm{He}} = X_{\mathrm{H_2}} \,\frac{X_{\mathrm{He}}}{X_{\mathrm{H_2}}},
\end{equation}
where the He to H$_{2}$ ratio is calculated from \citet{Asplund_2021}
\begin{equation}
    \frac{X_{\mathrm{He}}}{X_{\mathrm{H_2}}} = 0.164.
\end{equation}

As an option, useful for ultra hot Jupiters, but not used in the current study, I extend the formalism to include the H/H$_{2}$ fraction to recover the effects of potential H$_{2}$ dissociation.
I first define an H/H$_{2}$ VMR ratio as a retrievable quantity
\begin{equation}
    f = \frac{X_{\mathrm{H}}}{X_{\mathrm{H_{2}}}},
\end{equation}
the remainder background fraction
\begin{equation}
    X_{\mathrm{bg}} = 1-\sum_{i,\,i\neq \mathrm{He},\mathrm{H_2},\mathrm{H}}^{n} X_i,
\end{equation}
and He to H solar ratio
\begin{equation}
    \epsilon_{\rm He} = \frac{{\rm He}}{\rm {H}} = 0.082,
\end{equation}
taken from \citet{Asplund_2021}.
I then define $N_{\rm H}$ to be a dimensionless abundance of hydrogen nuclei in the filling mixture
\begin{equation}
    N_{\mathrm{H}} = \frac{X_{\mathrm{bg}}} {\frac{1+f}{2+f} + \epsilon_{\mathrm{He}}}.
\end{equation}
The H$_{2}$, H and He VMRs are then
\begin{equation}
X_{\mathrm{H_2}}=\frac{N_{\mathrm{H}}}{f+2},
\qquad
X_{\mathrm{H}}=f\,X_{\mathrm{H_2}},
\qquad
X_{\mathrm{He}}=\epsilon_{\mathrm{He}}\,N_{\mathrm{H}}.
\end{equation}
I note that in certain circumstances $X_{\mathrm{H}}$ $>$ $X_{\mathrm{H_2}}$, which can be taken into account when setting prior bounds for the H/H$_{2}$ ratio.
Combined with a simple electron number density fraction, $f_{\rm e^{-}}$, as a retrieved parameter via
\begin{equation}
    f_{\rm e^{-}} = \frac{n_{\rm e^{-}}}{n_{\rm tot}},
\end{equation}
where $n_{\rm e^{-}}$ [cm$^{-3}$] is the electron number density and $n_{\rm tot}$ [cm$^{-3}$] the background gas number density, this scheme enables a consistent way to recover free-free H$^{-}$ opacity, which can be an important continuum opacity source for ultra hot Jupiter atmospheres \citep[e.g.][]{Parmentier_2018}.

\section{1D Transmission Spectra}
\label{sec:trans}

For the transmission spectra calculations, I use the 1D geometric, absorption only limit of the path distribution method presented in \citet{MacDonald_2022} (Equation 19), which is based on the original Monte-Carlo formulation in \citet{Robinson_2017}.
This scheme enables a simple, fast method for calculating the transit limb, or slant, path.
Briefly, this method constructs a geometric, slant path matrix, $\mathcal{P}_{{\rm 1D}}$,  following
\begin{equation}
    \mathcal{P}_{{\rm 1D}, il} = 
    \begin{cases}
          0, &  r_{{\rm up}, l} \le b_{i}, \\
          \frac{2}{\Delta r_{l}}\left(\sqrt{r_{{\rm up}, l}^{2} - b_{i}^{2}}\right), &  r_{{\rm low}, l} < b_{i} < r_{{\rm up}, l}, \\ 
           \frac{2}{\Delta r_{l}}\left(\sqrt{r_{{\rm up}, l}^{2} - b_{i}^{2}} - \sqrt{r_{{\rm low}, l}^{2} - b_{i}^{2}}\right), &  r_{{\rm low}, l} \ge b_{i},
    \end{cases}
\end{equation}
where $r_{l}$ [cm] is the radius at layer $l$ with upper $r_{{\rm up}, l}$ [cm] and lower $r_{{\rm low}, l}$ [cm] level interfaces, and $b_{i}$ [cm] the impact parameter at layer $i$.
The transit limb optical depth can then be related to the vertical optical depth through matrix multiplication
\begin{equation}
    \tau_{\rm  trans} = \mathcal{P}_{{\rm 1D}} \cdot \Delta\tau_{\rm vert},
\end{equation}
which can then be used in the transmission spectra equation \citep[e.g.][]{Dobbs-Dixon_2013}.
The utility of this approach comes from the fact that $\mathcal{P}_{{\rm 1D}}$ is only required to be calculated once, and can be reused to find the slant optical path of any opacity component.
Generalisation of the path matrix to multiple dimensions is derived in \citet{MacDonald_2022}.

\section{1D Emission Spectra}
\label{sec:em}

For the emission spectra calculations, I use the 1D, linear-in-tau, $\delta$-M scaled, extended absorption approximation scheme ($\delta$-EAA) described in \citet{Lee_2024d}, taking the 8-stream Gauss–Jacobi angles and weights in \citet{Hogan_2024} for the intensity to flux integration. 
In \citet{Lee_2024d}, this method was found to produce comparable top of atmosphere fluxes to within 2\% of the Toon89 \citep{Toon_1989} and \textit{DISORT} \citep{Stamnes_1988} methods in cloudy hot Jupiter scenarios.

For hot Jupiter dayside emission retrievals, I follow the standard relative planetary and stellar flux calculation \citep[e.g.][]{Seager_2010}
\begin{equation}
    \frac{F_{\rm p}}{F_{\star}} = \frac{s F_{\rm toa}}{F_{\star}}\left(\frac{R_{\rm p}}{R_{\star}}\right)^{2},
\end{equation}
where $F_{\rm toa}$ [erg s$^{-1}$ cm$^{-2}$ cm$^{-1}$] is the wavelength dependent outgoing flux at the top of the atmosphere of the planet, $F_{\star}$ [erg s$^{-1}$ cm$^{-2}$ cm$^{-1}$] the stellar surface flux, $R_{\rm p}$ [cm] the radius of the planet at the reference pressure and $R_{\star}$ [cm] the stellar radius.
I also include an optional retrieved dilution factor, $s$ [0, 1], recommended by \citet{Taylor_2020} to mitigate potential geometric biases in a 1D retrieval setup.
I do not correct for wavelength dependent photospheric area emission examined by \citet{Fortney_2019} in \textit{Exo Skryer}.

For the isolated brown dwarf atmosphere scenarios, I remove the stellar spectrum dependence and scale to the distance of the brown dwarf
\begin{equation}
    F_{\rm bd} = sF_{\rm toa} \left(\frac{R_{\rm bd}}{D}\right)^{2}
\end{equation}
where $D$ [cm] is the distance to the brown dwarf, $s$ [0,1] remains a geometric bias factor, while the radius of the brown dwarf is derived from 
\begin{equation}
    R_{\rm bd} = \sqrt{\frac{G M_{\rm bd}}{g}},
\end{equation}
where the $M_{\rm bd}$ [g] and gravity $g$ [cm s$^{-2}$] are the retrieved quantities inside \textit{Exo Skryer}.
This is similar to the approach used in \citet{Kitzmann_2020}, who used a scaling parameter, $f$, which acts as a nuisance scaling parameter representing errors in the inhomogeneity, distance and mass of the object.

\section{Retrieval Model Examples}
\label{sec:ret}

\begin{table*}[]
    \centering
    \caption{Computational performance details for each experiment. All experiments used 1000 live points with a termination condition of $\Delta \ln Z$ = 0.1.}
    \begin{tabular}{c|c|c|c|c|c}
        Experiment & Opacity Scheme & Sampler \& computation & No. wavelengths \& range & No. parameters & Total runtime (days, hrs, mins)  \\ \hline \hline 
         HD 209458b (Trans.)  & OS (R = 20,000) & \textit{dynesty}: 1 CPU + 1 GPU & 40,471: 0.3-5.05 $\mu$m & 11 & 1h 15m\\
         WASP-107b (Trans.)   & OS (R = 20,000) & \textit{dynesty}: 1 CPU + 1 GPU  & 62,217: 0.6-13.7 $\mu$m & 28 & 2d 7h 53m \\
         WASP-17b  (Trans.)  & OS (R = 20,000) & \textit{dynesty}: 1 CPU + 1 GPU & 71,193: 0.3-12.0 $\mu$m & 43 & 1d 16h 15m \\
         HD 189733b  (Em.)  &  OS (R = 20,000) & \textit{dynesty}: 1 CPU + 1 GPU & 40,018: 1.1-12.0 $\mu$m & 44 & 1d 13h 31m\\
         Gliese 229 B (Em.)   &  OS (R = 40,000) & \textit{dynesty}: 1 CPU + 1 GPU & 54240: 1.0-5.1 $\mu$m & 15 & 9h 8m \\ \hline \hline     
    \end{tabular}
    \label{tab:ret}
\end{table*}

In this section, I apply \textit{Exo Skryer} across a variety of contemporary retrieval scenarios already published in the literature, or typical scenarios encountered by practitioners in the field.
This allows a benchmark of the \textit{Exo Skryer} framework to check that it is producing consistent results to other models.
I also state the start-to-end completion time of each model.
For brevity, I primarily focus on parameter estimation and not Bayesian evidence or model comparison for each scenario.
In Table. \ref{tab:ret}, I provide an overview of the retrieval setup, as well as the end-to-end runtime for each experiment.

\subsection{Transmission - HD 209458b}
\label{sec:HD209}

\begin{figure*}
    \centering
    \includegraphics[width=\linewidth]{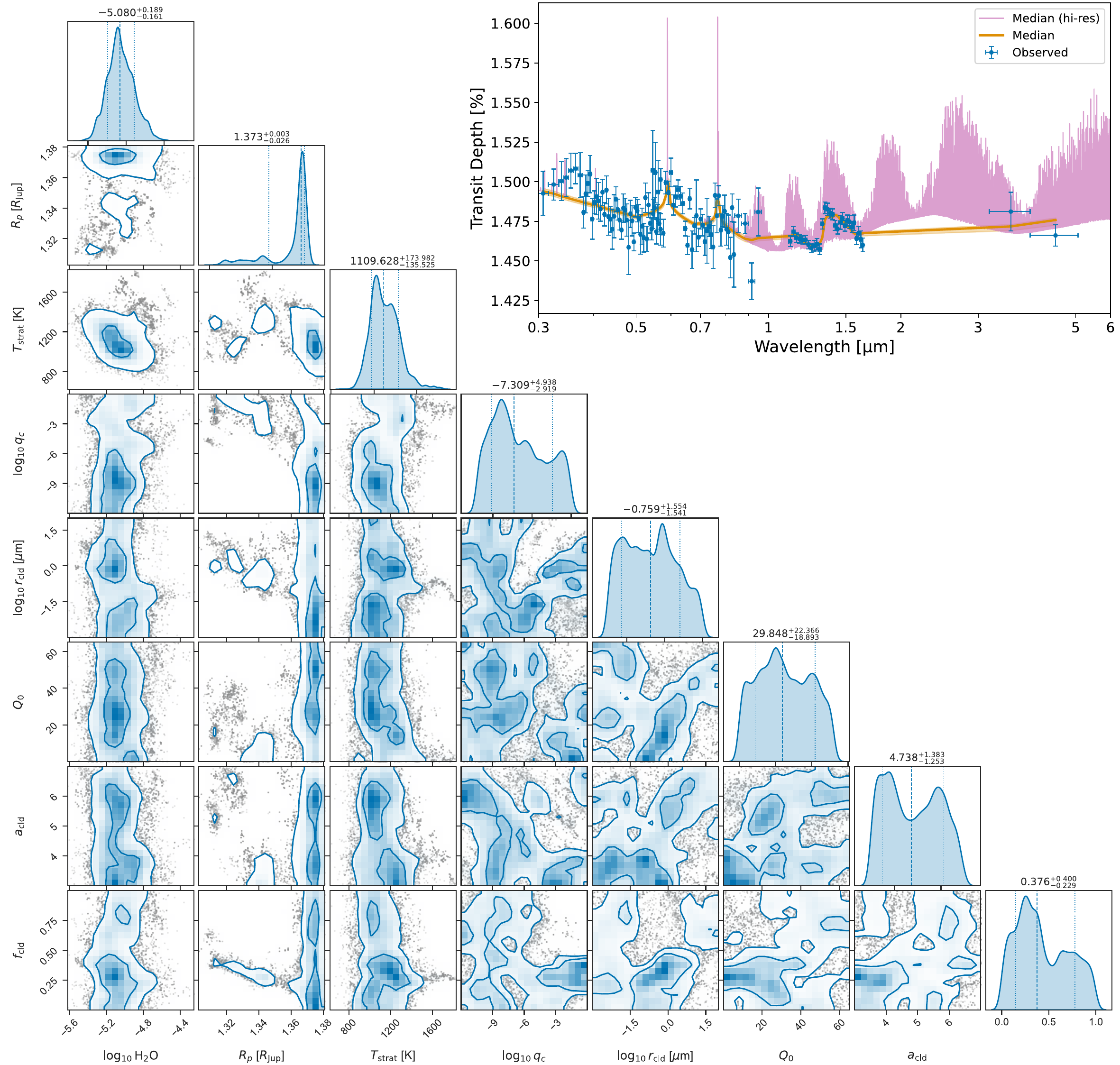}
    \caption{Equal weight posterior samples and kernel density estimation for the HD 209458b pre-JWST transmission spectra.
    The median and 1$\sigma$ confidence intervals are shown as the dashed and dotted lines respectively.
    The plot in the top right shows the best fit median spectra (orange) with 1$\sigma$ (shaded) region to the observed HD 209458b pre-JWST transmission spectra data (blue). 
    The high-resolution median spectrum is in purple.}
    \label{fig:HD209_dia}
\end{figure*}

\begin{table}[]
    \centering
    \caption{HD 209458b pre-JWST transmission spectrum retrieval setup and results.}    
    \begin{tabular}{c|l|c} \hline \hline
        Parameter & Prior  & Posterior \\ \hline
         $R_{\star}$ [$R_{\odot}$]&  $\delta$(1.187) & - \\
         $p_{\rm bot}$ [bar]&  $\delta$(100) & - \\
         $p_{\rm top}$ [bar]&  $\delta$(10$^{-8}$) & - \\
         $p_{\rm ref}$ [bar]&   $\delta$(10)  & - \\
         $\rho_{\rm d}$ [g cm$^{-3}$]&   $\delta$(2.5)  & - \\
         $Q_{1}$ [-]&   $\delta$(4.21)  & - \\
         $\log_{10} g_{\rm ref}$ [cm s$^{-2}$]&   $\mathcal{U}$(2.7, 3.3)  & 3.19$^{+0.06}_{-0.08}$\\
         $R_{\rm p, ref}$ [$R_{\rm jup}$]&  $\mathcal{U}$(1.1, 1.5) & 1.37$^{+0.00}_{-0.03}$  \\
         $T_{\rm strat}$ [K] &  $\mathcal{U}$(100, 3000) &  1110$^{+174}_{-136}$ \\
         $\log_{10}$ H$_{2}$O [-]& $\mathcal{U}$(-12, -2)  & -5.08$^{+0.19}_{-0.16}$ \\
         $\log_{10}$ Na [-]&  $\mathcal{U}$(-12, -2) &  -6.65$^{+0.25}_{-0.29}$ \\
         $\log_{10}$ K [-]& $\mathcal{U}$(-12, -2)  &  -7.96$^{+0.29}_{-0.35}$  \\
         $\log_{10}$ $q_{\rm c}$ [g g$^{-1}$]&  $\mathcal{U}$(-12, 0) &  -7.31$^{+4.94}_{-2.92}$ \\
         $\log_{10}$ $r_{\rm cld}$ [$\mu$m]&  $\mathcal{U}$(-3, 2) &  -0.76$^{+1.55}_{-1.54}$ \\
         $Q_{0}$ [-]& $\mathcal{U}$(0.1, 65) & 29.85$^{+22.37}_{-18.89}$ \\
         $a_{\rm cld}$ [-]& $\mathcal{U}$(3, 7)  & 4.74$^{+1.38}_{-1.25}$ \\
         $f_{\rm cld}$ [-]&  $\mathcal{U}$(0, 1) &  0.38$^{+0.40}_{-0.23}$ \\ \hline \hline
    \end{tabular} 
    \label{tab:HD209b}
\end{table}

In an initial test of the \textit{Exo Skryer} retrieval framework, I follow an HD 209458b transmission spectrum retrieval setup presented in \citet{Barstow_2020}.
\citet{Barstow_2020} used HST+Spitzer data from \citet{Sing_2016}, retrieving H$_{2}$O, Na, K mixing ratios, as well as cloud parameters.
I use a similar setup to \citet{Barstow_2020}, namely, retrieving the same molecular species and using the stratosphere + adiabatic $T$-$p$ parameterisation in \citet{Barstow_2020}.
I use the modified \citet{Fisher_2018} cloud opacity scheme (Section \ref{sec:F18}) with a constant condensed mass mixing ratio vertical profile, which in aggregate is highly similar to \citet{Barstow_2020}.
Using \textit{dynesty} with 11 sampled parameters (Table \ref{tab:HD209b}), the elapsed time from start to finish of the retrieval model (OS at $R$ = 20,000) was 1 hour 15 minutes.

In Figure \ref{fig:HD209_dia}, I present the posterior distribution of the retrieval and best-fit median spectra.
In comparing the posterior results to the retrievals performed with \textit{NEMESIS} in \citet{Barstow_2020} (Figure 9), \textit{Exo Skryer} is producing consistent results with respect to the H$_{2}$O mixing ratio, temperatures and reference radius.
However, the main cloud opacity driver parameter, $k_{0}$ in \citet{Barstow_2020} and $q_{\rm c}$ here remains quite unconstrained, with several orders of magnitude between the median and 1$\sigma$ confidence intervals, with bimodal structures present across the cloud parameters.
This bimodal structure suggests that a large fraction of cloud mass contained in small, $\sim$0.01 $\mu$m sized particles or a smaller mass mixing ratio with large $\sim$1 $\mu$m particles can reasonably fit the data.
$Q_{0}$ and the scattering index, $a_{\rm cld}$, are consistent with \citet{Barstow_2020}, but \textit{Exo Skryer} fits a smaller particle size and larger cloud fraction median pairing ($r_{\rm cld}$ = 0.17 $\mu$m, $f_{\rm cld}$ = 0.38) than \citet{Barstow_2020} ($r_{\rm eff}$ = 3.39 $\mu$m, $f_{\rm cld}$ = 0.33), though \textit{Exo Skryer} has larger 1$\sigma$ ranges than those in \citet{Barstow_2020}.
Overall, the physical interpretation conclusions from this retrieval effort are highly similar to \citet{Barstow_2020}, in that H$_{2}$O can be reasonably constrained to within $\approx$0.1 dex, but cloud structure information remains highly uncertain within the pre-JWST data.

\subsection{Transmission - WASP-107b}
\label{sec:W107b}

\begin{figure*}
    \centering
    \includegraphics[width=\linewidth]{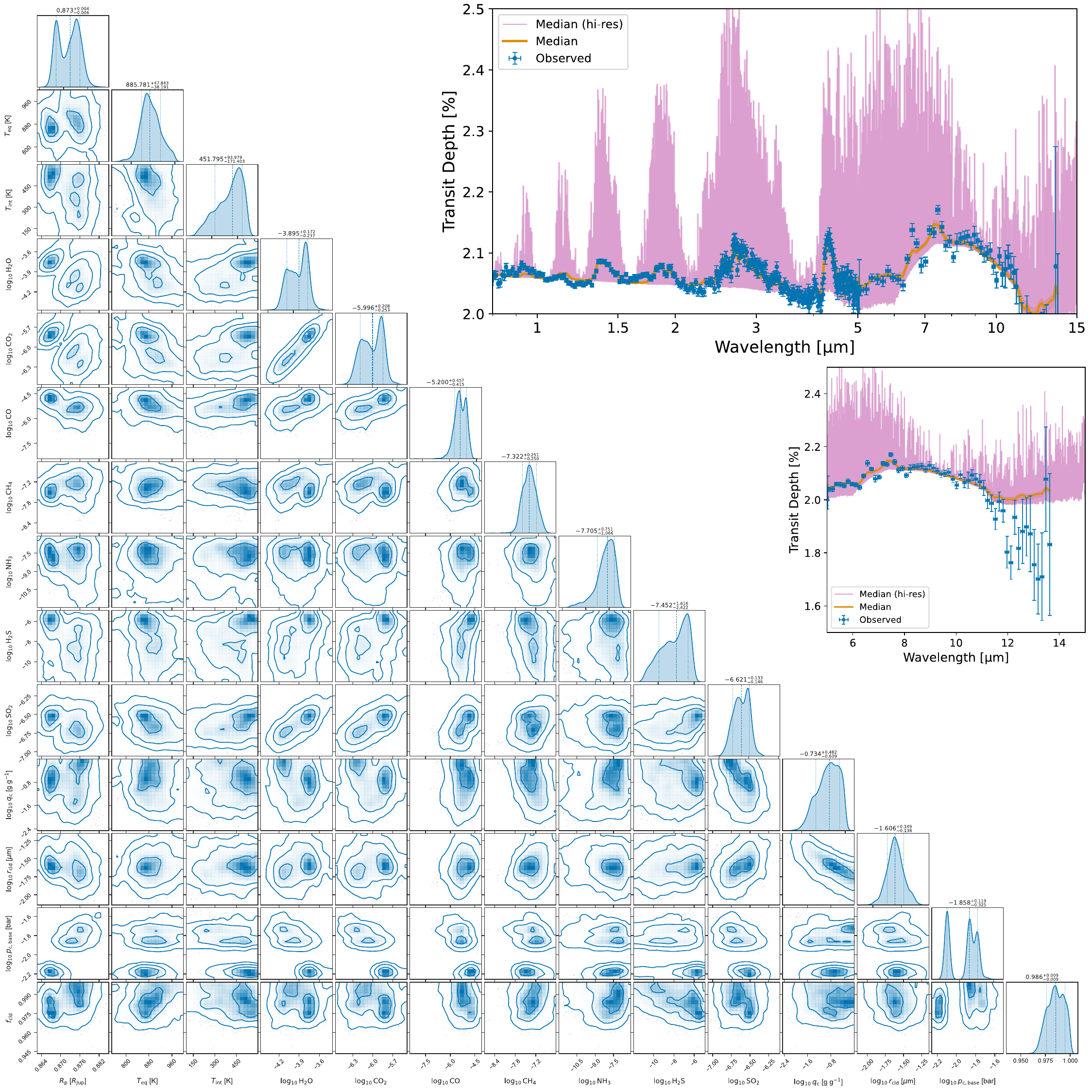}
    \caption{Equal weight posterior samples and kernel density estimation for the WASP-107b JWST transmission spectra.
    The median and 1$\sigma$ confidence intervals are shown as the dashed and dotted lines respectively.
    The top left plots show the observational data (blue points), median model with 1$\sigma$ ranges (orange) and high-resolution median model (purple).}
    \label{fig:W107b}
\end{figure*}

\begin{table}[]
    \centering
    \caption{WASP-107b JWST transmission spectrum retrieval setup and results.}    
    \begin{tabular}{c|c|c} \hline \hline
        Parameter & Prior  & Posterior \\ \hline
         $R_{\star}$ [$R_{\odot}$]& $\delta$(0.67)& - \\
         $p_{\rm bot}$ [bar]&  $\delta$(1000) & - \\
         $p_{\rm top}$ [bar]&  $\delta$(10$^{-8}$) & - \\
         $p_{\rm ref}$ [bar]&   $\delta$(10)  & - \\
         $\rho_{\rm d}$ [g cm$^{-3}$]&  $\delta$(2.5) & - \\
         $Q_{1}$ [-]&  $\delta$(4.21) & - \\
         $f_{\rm hem}$ [-]&  $\delta$(0.25) & - \\
         $x_{0}$ [-]&  $\delta$(0.5) & - \\
         $\log_{10} g$ [cm s$^{-2}$]&  $\mathcal{U}$(2.0, 3.5) & 2.72$^{+0.02}_{-0.02}$\\
         $R_{\rm p, ref}$ [$R_{\rm jup}$]&  $\mathcal{U}$(0.66, 1.22) & 0.87$^{+0.00}_{-0.01}$\\
         $T_{\rm eq}$ [K] &  $\mathcal{U}$(500, 1000) &  886$^{+48}_{-38}$\\
         $T_{\rm int}$ [K] &  $\mathcal{U}$(100, 600) &  451$^{+94}_{-168}$\\
         $\log_{10}$ $\kappa_{\rm ir}$ [cm$^{2}$ g$^{-1}$]& $\mathcal{U}$(-6, 2)  & -4.74$^{+0.35}_{-0.36}$\\
         $\log_{10}$ $\gamma_{\rm v}$ [-]& $\mathcal{U}$(-4, 2)  & -1.57$^{+0.87}_{-1.27}$\\
         $\log_{10}$ $p_{\rm t}$ [bar] & $\mathcal{U}$(-8, 3)  & -6.21$^{+2.07}_{-1.27}$\\
         $q_{\rm irr}$ [-]& $\mathcal{U}$(0, 0.66)  & 0.36$^{+0.15}_{-0.17}$\\
         $\beta$ [-]& $\mathcal{U}$(0, 1)  & 0.30$^{+0.39}_{-0.25}$\\
         $\log_{10}$ H$_{2}$O [-]& $\mathcal{U}$(-12, -2)  & -3.89$^{+0.17}_{-0.24}$\\
         $\log_{10}$ CO$_{2}$ [-]& $\mathcal{U}$(-12, -2)  & -6.00$^{+0.21}_{-0.25}$\\
         $\log_{10}$ CO [-]& $\mathcal{U}$(-12, -2)  &  -5.19$^{+0.44}_{-0.41}$\\
         $\log_{10}$ CH$_{4}$ [-]& $\mathcal{U}$(-12, -2)  & -7.32$^{+0.25}_{-0.26}$\\
         $\log_{10}$ NH$_{3}$ [-]&  $\mathcal{U}$(-12, -2) & -7.70$^{+0.75}_{-1.06}$ \\
         $\log_{10}$ H$_{2}$S [-]& $\mathcal{U}$(-12, -2)  & -7.48$^{+1.62}_{-2.46}$\\
         $\log_{10}$ SO$_{2}$ [-]& $\mathcal{U}$(-12, -2)  & -6.62$^{+0.13}_{-0.15}$ \\
         $\log_{10}$ $r_{\rm cld}$ [$\mu$m] & $\mathcal{U}$(-3, 2)  & -1.61$^{+0.15}_{-0.13}$\\
         $\log_{10}$ $q_{\rm c, base}$ [g g$^{-1}$] & $\mathcal{U}$(-12, 0) & -0.72$^{+0.47}_{-0.63}$\\
         $\log_{10}$ $p_{\rm c, base}$ [bar] & $\mathcal{U}$(-8, 3)  & -1.86$^{+0.12}_{-0.32}$ \\
         $\log_{10}$ $\alpha_{\rm cld}$ [-] & $\mathcal{U}$(-2, 2)  & 0.77$^{+0.05}_{-0.08}$ \\
         $Q_{0}$ [-] &  $\mathcal{U}$(0.1, 65) & 27$^{+11}_{-7}$ \\
         $a_{\rm cld}$ [-] &  $\mathcal{U}$(2, 5) & 3.24$^{+0.23}_{-0.15}$ \\
         $A$ [-] &  $\mathcal{U}$(0, 10) & 6.82$^{+1.90}_{-2.31}$ \\
         $\lambda_{0}$ [$\mu$m] &  $\mathcal{U}$(7, 10) & 7.48$^{+0.13}_{-0.10}$ \\
         $\omega$ [$\mu$m] &  $\mathcal{U}$(0, 4) & 0.68$^{+0.05}_{-0.05}$ \\
         $\xi$ [-] &  $\mathcal{U}$(0, 10) & 2.32$^{+0.36}_{-0.39}$ \\
         $f_{\rm cld}$ [-] &  $\mathcal{U}$(0, 1) & 0.99$^{+0.01}_{-0.01}$ \\
         $\Delta_{\rm MIRI}$ [ppm] &  $\mathcal{U}$(-50, 500) &  321$^{+43}_{-46}$\\
         \hline \hline
    \end{tabular} 
    \label{tab:W107b}
\end{table}

WASP-107b \citep{Anderson_2017} is a warm-Neptune exoplanet, recently characterised by HST \citep{Kreidberg_2018, Spake_2018} and JWST \citep{Sing_2024, Welbanks_2024, Dyrek_2024, Krishnamurthy_2026}.
For the retrieval setup, I use the combined JWST NIRCam + JWST MIRI data from \citet{Welbanks_2024}, and the JWST NIRISS SOSS data from \citet{Krishnamurthy_2026} following closely the free retrieval setup from \citet{Welbanks_2024}, as well as retrieving an offset for the MIRI data as in \citet{Welbanks_2024}.
I use the combined parameterised continuum cloud opacity scheme and mid-infrared feature described in Section \ref{sec:feat}.
Using \textit{dynesty} with 28 parameters (Table \ref{tab:W107b}) the retrieval took 2 days 7 hours 53 minutes (OS at $R$ = 20,000).

Figure \ref {fig:W107b} presents the posterior distribution and best-fit median spectra for the WASP-107b transmission spectra retrieval.
I recover a very different free-retrieval outcome to that found in \citet{Welbanks_2024}, whereby I recover 1-2 dex lower VMR for all gas species. 
This is possibly a result of the assumed cloud structures and opacity, with \citet{Welbanks_2024} using a grey deck and powerlaw approach, while here I use a non-grey, vertically varying mass mixing ratio scheme. 
The wavelength dependent approach used here allows the strong muting of the H$_{2}$O features at near-IR wavelengths, but has a reduced cloud opacity at the longer wavelengths, enabling a smaller mixing ratio to recover the required amplitude of the absorption features. 
In contrast to the enhanced metallicity scenario recovered by \citet{Welbanks_2024}, the results suggest a thick, upper atmosphere cloud component which only allows the strongest lines to be probed in transmission, recovering a depleted atmosphere, driven by the missing information content provided by spectral windows for each species. 
The post-processed transmission spectra are reminiscent of studies that performed high-resolution modelling with strong upper atmosphere cloud opacity \citep[e.g.][]{Gandhi_2020}.

\subsection{Transmission - WASP-17b}
\label{sec:W17b}

\begin{figure*}
    \centering
    \includegraphics[width=\linewidth]{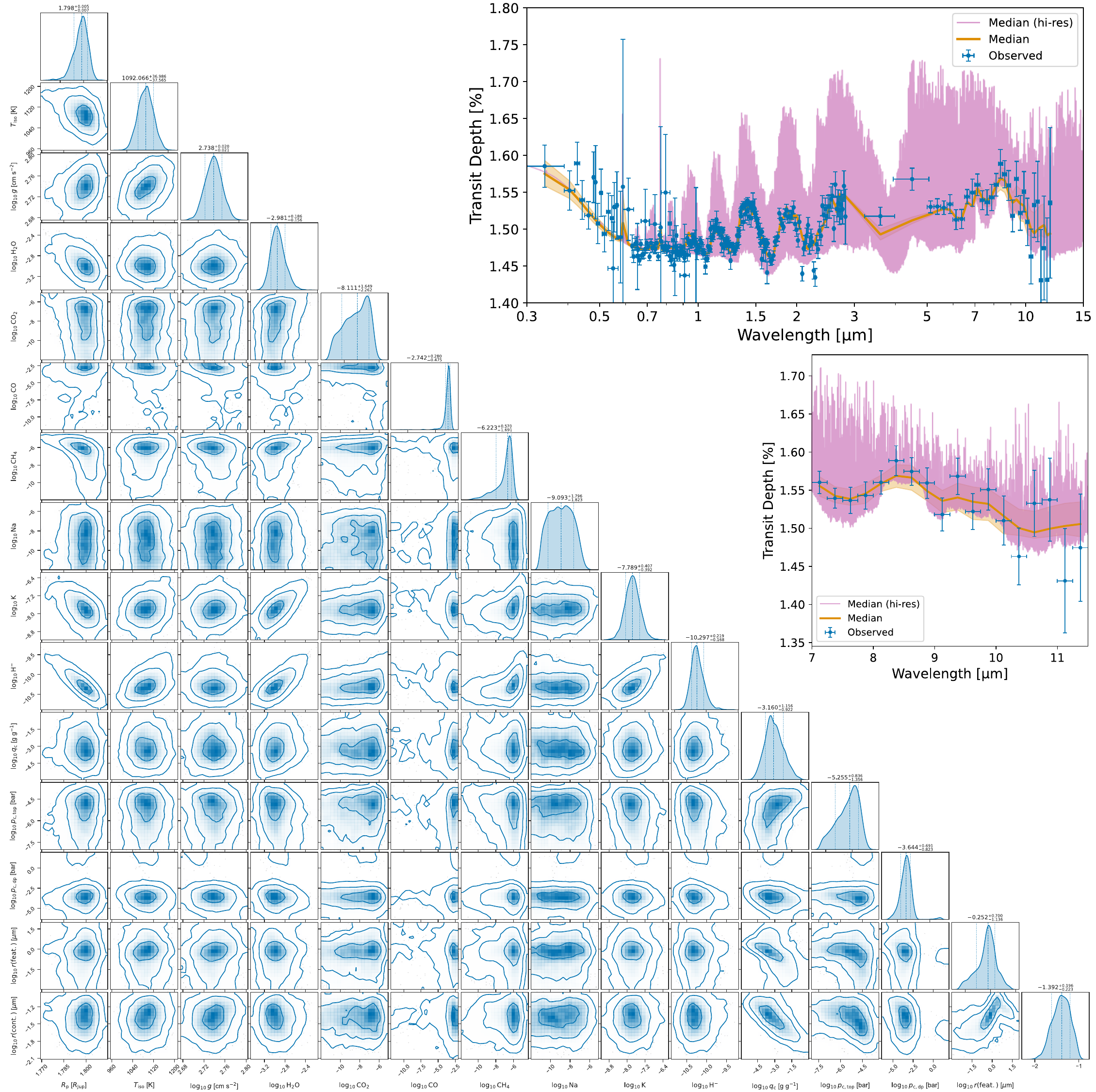}
    \caption{Equal weight posterior samples and kernel density estimation for the WASP-17b HST and JWST combined transmission spectra retrieval from \citet{Louie_2025}.
    The median and 1$\sigma$ confidence intervals are shown as the dashed and dotted lines respectively.
    The plots in the top right show the best fit median spectra (orange) with 1$\sigma$ (shaded) region to the observed transmission spectra data (blue), taken from \citet{Louie_2025}, for the full spectral range (top right) and MIRI LRS wavelength range (middle right). 
    The high-resolution median spectrum is in purple.}
    \label{fig:W17b}
\end{figure*}

\begin{figure}
    \centering
    \includegraphics[width=\linewidth]{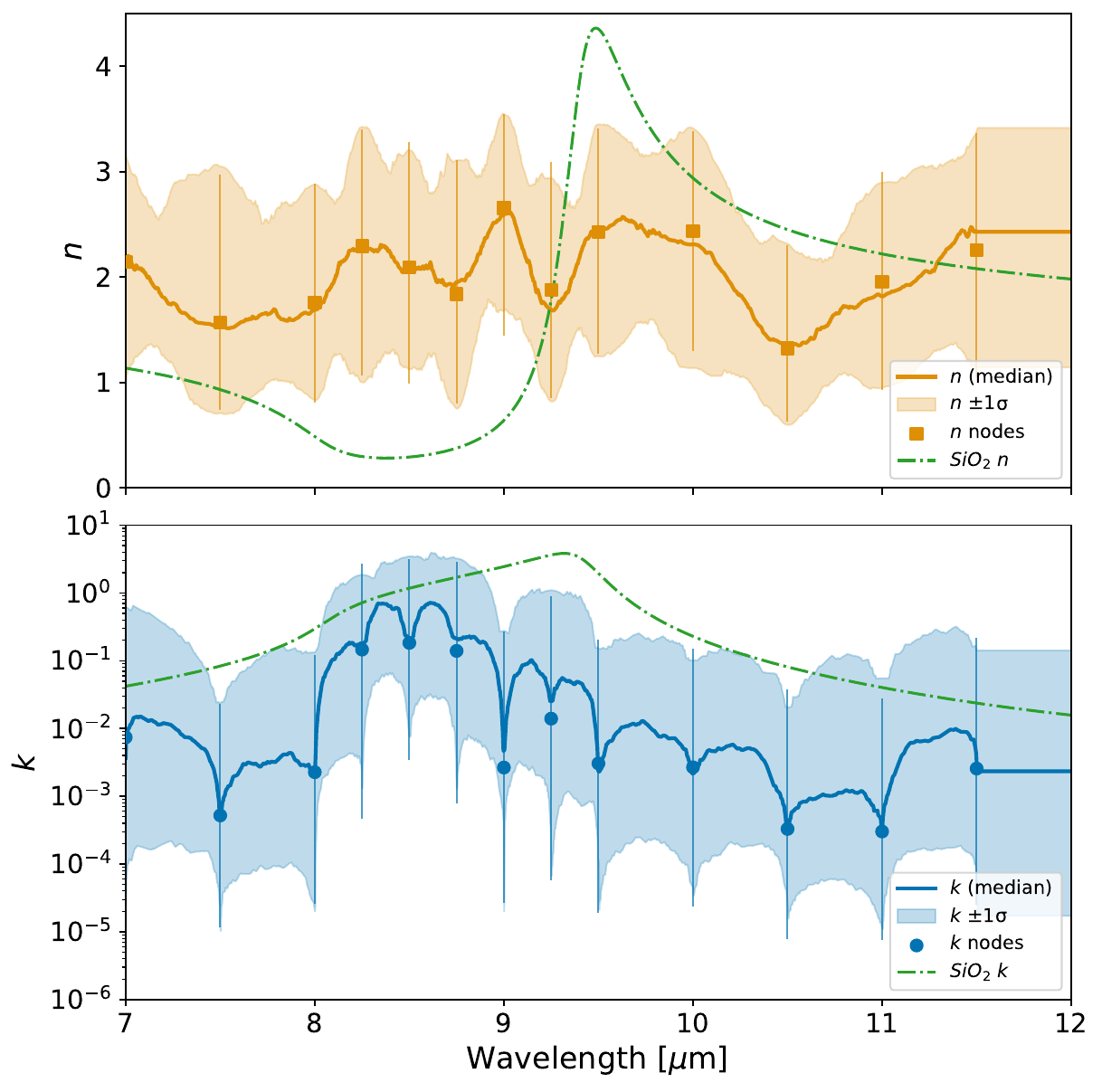}
    \caption{Direct $n$ (top, orange) and $k$ (bottom, blue) retrieval results from the WASP-17b transmission spectrum test.
    The glassy $\alpha$-SiO$_{2}$ optical constants from \citet{Zeidler_2013} are plotted with the dash-dot green lines.}
    \label{fig:W17b_nk}
\end{figure}

\begin{table}[]
    \centering
    \caption{WASP-17b JWST transmission spectrum retrieval setup and results.}    
    \begin{tabular}{c|c|c} \hline \hline
        Parameter & Prior  & Posterior \\ \hline
         $R_{\star}$ [$R_{\odot}$]& $\delta$(1.583)& - \\
         $p_{\rm bot}$ [bar]&  $\delta$(100) & - \\
         $p_{\rm top}$ [bar]&  $\delta$(10$^{-8}$) & - \\
         $p_{\rm ref}$ [bar]&   $\delta$(10)  & - \\
         $\rho_{\rm d}$ [g cm$^{-3}$]&  $\delta$(2.5) & - \\
         $Q_{1}$ [-]&  $\delta$(4.21) & - \\
         $\log_{10} g$ [cm s$^{-2}$]&  $\mathcal{U}$(2.2, 3.3) &  2.74$^{+0.02}_{-0.02}$\\
         $R_{\rm p, ref}$ [$R_{\rm jup}$]&  $\mathcal{U}$(1.5895, 2.1505) & 1.80$^{+0.00}_{-0.01}$ \\
         $T_{\rm iso}$ [K] &  $\mathcal{U}$(400, 2300) &  1092$^{+37}_{-37}$\\
         $\log_{10}$ H$_{2}$O [-]& $\mathcal{U}$(-12, -2)  & -2.98$^{+0.18}_{-0.15}$ \\
         $\log_{10}$ CO$_{2}$ [-]& $\mathcal{U}$(-12, -2)  & -8.15$^{+1.68}_{-2.23}$ \\
         $\log_{10}$ CO [-]& $\mathcal{U}$(-12, -2)  & -2.74$^{+0.28}_{-0.49}$ \\
         $\log_{10}$ CH$_{4}$ [-]& $\mathcal{U}$(-12, -2)  &  -6.22$^{+0.56}_{-1.70}$\\
         $\log_{10}$ Na [-]&  $\mathcal{U}$(-12, -2) & -9.13$^{+1.82}_{-1.80}$ \\
         $\log_{10}$ K [-]& $\mathcal{U}$(-12, -2)  & -7.79$^{+0.41}_{-0.41}$ \\
         $\log_{10}$ H$^{-}$ [-]& $\mathcal{U}$(-12, -2)  & -10.64$^{+0.10}_{-0.11}$ \\
         $Q_{0}$ [-] &  $\mathcal{U}$(0.1, 65) & 36$^{+20}_{-20}$ \\
         $a_{\rm cld}$ [-] &  $\mathcal{U}$(3, 7) & 6.73$^{+0.20}_{-0.34}$ \\         
         $\log_{10}$ $r_{\rm F18}$ [$\mu$m] & $\mathcal{U}$(-3, 2)  & -1.39$^{+0.20}_{-0.22}$\\
         $\log_{10}$ $q_{\rm c}$ [g g$^{-1}$] & $\mathcal{U}$(-12, 0)  & -3.17$^{+1.17}_{-0.95}$ \\
         $\log_{10}$ $p_{\rm c, top}$ [bar] & $\mathcal{U}$(-8, 2)  & -5.26$^{+0.84}_{-1.35}$  \\
         $\log_{10}$ $p_{\rm c, dp}$ [bar] & $\mathcal{U}$(-8, 2)  &  -3.63$^{+0.68}_{-0.81}$ \\
         $\log_{10}$ $r_{\rm nk}$ [$\mu$m] & $\mathcal{U}$(-3, 2)  & -0.27$^{+0.70}_{-1.14}$\\
         $n_{0...12}$ [-] & $\mathcal{U}$(0.3, 4.0)  & Figure \ref{fig:W17b_nk} \\
         $\log_{10}$ $k_{0...12}$ [-] & $\mathcal{U}$(-6, 1)  & Figure \ref{fig:W17b_nk} \\         
         \hline \hline
    \end{tabular} 
    \label{tab:W17b}
\end{table}

To test the proposed direct optical constant retrieval scheme, I perform a similar retrieval setup to that performed for WASP-17b in \citet{Louie_2025}, who published new SOSS data and combined them with the previous HST STIS+WFC3 \citep{Alderson_2022} and MIRI LRS \citep{Grant_2023} data.
I follow the enhanced `B' scheme from \citet{Louie_2025}, which included retrieval of H$_{2}$O, CO, CO$_{2}$, CH$_{4}$, Na, K and H$^{-}$ (bound-free).
I follow similar priors and setup to the \textit{POSEIDON} \citep{MacDonald_2023} framework used in \citet{Louie_2025}, assuming an isothermal 100 layer atmosphere evenly log-spaced between 100 and 10$^{-8}$ bar.
I include Rayleigh scattering from H$_{2}$ and He as well as CIA from H$_{2}$-H$_{2}$ and H$_{2}$-He pairs.
For the cloud model, I use the same slab profile approach used in \textit{POSEIDON}, but only retrieve the cloud properties in the 7-11.5 $\mu$m wavelength range in order to focus on retrieving the $n$ and $k$ constants for the cloud absorption features.
The spectrum is therefore cloud free outside this wavelength range.
I take the HST + SOSS + Spitzer + MIRI spectrum presented in \citet{Louie_2025}, adding an offset to the MIRI data as retrieved in \citet{Louie_2025}.
Using \textit{dynesty} with 43 sampled parameters (Table \ref{tab:W17b}), the elapsed time from start to finish of the retrieval model was 1 day, 16 hours and 15 minutes for the $R$ = 20,000 model.

Figure \ref{fig:W17b} presents the posterior and best-fit median model compared to the observational data.
The retrieval produces similar results to \citet{Louie_2025}, with the main difference being a couple of magnitudes larger CO retrieved here compared to \citet{Louie_2025}.
This is probably driven by the lack of a cloud opacity continuum, requiring a larger CO opacity to compensate for the large Spitzer 4.5$\mu$m observational point.

In addition, Figure \ref{fig:W17b} also contains a zoom in on the 7-12 $\mu$m region, where the ($n$, $k$) optical constants were retrieved.
Here, the absorption feature around 8.7 $\mu$m is well described by the retrieved parameters, and the absorption peak shifted slightly blueward compared to the SiO$_{2}$ quartz retrieval models in \citet{Louie_2025}.
A secondary peak is seen at around 9.5 $\mu$m, suggesting the presence of a secondary cloud absorption band component, though the 1$\sigma$ confidence region is wider in this wavelength regime.
The retrieval results suggest a mass mixing ratio of $q_{\rm c}$ $\sim$ 10$^{-3.2}$, which is in line with expectations of the Si abundance at around 10-100x Solar metallicity \citep{Asplund_2021}.
I recover different particle sizes for the continuum, $r$ $\sim$ 0.04 $\mu$m, which allows a strong optical slope and muting at near-IR wavelengths, while the direct $n$-$k$ scheme recovers $r$ $\sim$ 0.54 $\mu$m, for the same vertical cloud mass mixing ratio structure.
This suggests a multi-modal or polydisperse size-distribution solution for the cloud particles may be present at high altitudes in the planet.

In Figure \ref{fig:W17b_nk}, I compare the retrieved optical constant to the glassy $\alpha$-SiO$_{2}$ quartz constants from \citet{Zeidler_2013} commonly used in the literature.
Though large uncertainties remain, for the $k$ constant, the model suggests a material that may be consistent with SiO$_{2}$, but with a slightly broader, bluer bump.
This conforms with the fitting of this feature to various SiO$_{2}$ polymorphs in \citet{Moran_2024}, who found the absorption peak of the SiO$_{2}$ can shift dependent on the crystalline state of the mineral.
The scattering component of the opacity is not dominant in this wavelength range, which heavily limits the ability of the retrieval to produce constraints on the $n$ constant.
Overall, this discrepancy between the retrieved $k$ and the database SiO$_{2}$ values warrants further detailed investigations into the material compositions of the cloud particles in the WASP-17b atmosphere.

\subsection{Emission - HD 189733 b}
\label{sec:HD189b}

\begin{table}[]
    \centering
    \caption{HD 189733b JWST emission spectrum retrieval setup and results.}    
    \begin{tabular}{c|c|c} \hline \hline
        Parameter & Prior  & Posterior \\ \hline
         $R_{\star}$ [$R_{\odot}$]& $\delta$(0.765)& - \\
         $p_{\rm bot}$ [bar]&  $\delta$(100) & - \\
         $p_{\rm top}$ [bar]&  $\delta$(10$^{-8}$) & - \\
         $p_{\rm ref}$ [bar]&   $\delta$(10)  & - \\
         $\rho_{\rm d}$ [g cm$^{-3}$]&  $\delta$(2.5) & - \\
         $f_{\rm hem}$ [-]&  $\delta$(0.25) & - \\
         $\log_{10} g$ [cm s$^{-2}$]&  $\mathcal{U}$(2.7, 3.3) & 3.02$^{+0.15}_{-0.15}$ \\
         $R_{\rm p, ref}$ [$R_{\rm jup}$]&  $\mathcal{U}$(1.0, 1.2) & 1.12$^{+0.03}_{-0.02}$ \\
         $T_{\rm eq}$ [K] &  $\mathcal{U}$(500, 2000) &  1634$^{+34}_{-38}$\\
         $T_{\rm int}$ [K] &  $\mathcal{U}$(100, 600) & 385$^{+130}_{-142}$ \\
         $s$ [-] & $\mathcal{U}$(0, 1) & 0.93$^{+0.04}_{-0.05}$ \\
         $\log_{10} \kappa_{\rm ir}$ [cm$^{2}$ g$^{-1}$] &  $\mathcal{U}$(-6, 2) &  -3.54$^{+0.32}_{-0.44}$ \\
         $\log_{10} \gamma_{\rm v}$ [-] &  $\mathcal{U}$(-3, 2) & -1.15$^{+0.30}_{-0.29}$ \\
         $q_{\rm irr, 0}$ [-] &  $\mathcal{U}$(0, 0.66) &  0.20$^{+0.05}_{-0.06}$ \\
         $\log_{10} p_{\rm t}$ [bar] &  $\mathcal{U}$(-7, 1) & -1.88$^{+0.19}_{-0.22}$ \\
         $\beta$ [-] &  $\mathcal{U}$(0, 1) & 0.65$^{+0.16}_{-0.15}$ \\
         $\log_{10}$ H$_{2}$O [-]& $\mathcal{U}$(-12, -2)  &  -2.69$^{+0.16}_{-0.18}$ \\
         $\log_{10}$ CO$_{2}$ [-]& $\mathcal{U}$(-12, -2)  & -5.42$^{+0.15}_{-0.16}$ \\
         $\log_{10}$ CO [-]& $\mathcal{U}$(-12, -2)  & -2.26$^{+0.18}_{-0.21}$ \\
         $\log_{10}$ CH$_{4}$ [-]& $\mathcal{U}$(-12, -2)  &  -7.43$^{+0.99}_{-2.07}$ \\
         $\log_{10}$ $r_{\rm cld}$ [$\mu$m] & $\mathcal{U}$(-3, 2)  & -1.78$^{+0.83}_{-0.80}$ \\
         $\log_{10}$ $q_{\rm c, base}$ [g g$^{-1}$] & $\mathcal{U}$(-12, 1)  & -4.09$^{+0.83}_{-0.50}$ \\
         $\log_{10}$ $p_{\rm c, base}$ [bar] & $\mathcal{U}$(-8, 2)  & -0.79$^{+1.49}_{-1.32}$ \\
         $\log_{10}$ $\alpha_{\rm c}$ [-] & $\mathcal{U}$(-2, 2)  &  -0.77$^{+0.56}_{-0.60}$ \\ 
         $n_{0...12}$ [-] & $\mathcal{U}$(0.3, 4.0)  &  Figure \ref{fig:HD189b_nk}\\
         $\log_{10}$ $k_{0...12}$ [-] & $\mathcal{U}$(-6, 1)  &  Figure \ref{fig:HD189b_nk} \\
         \hline \hline
    \end{tabular} 
    \label{tab:HD189b}
\end{table}

\begin{figure*}
    \centering
    \includegraphics[width=\linewidth]{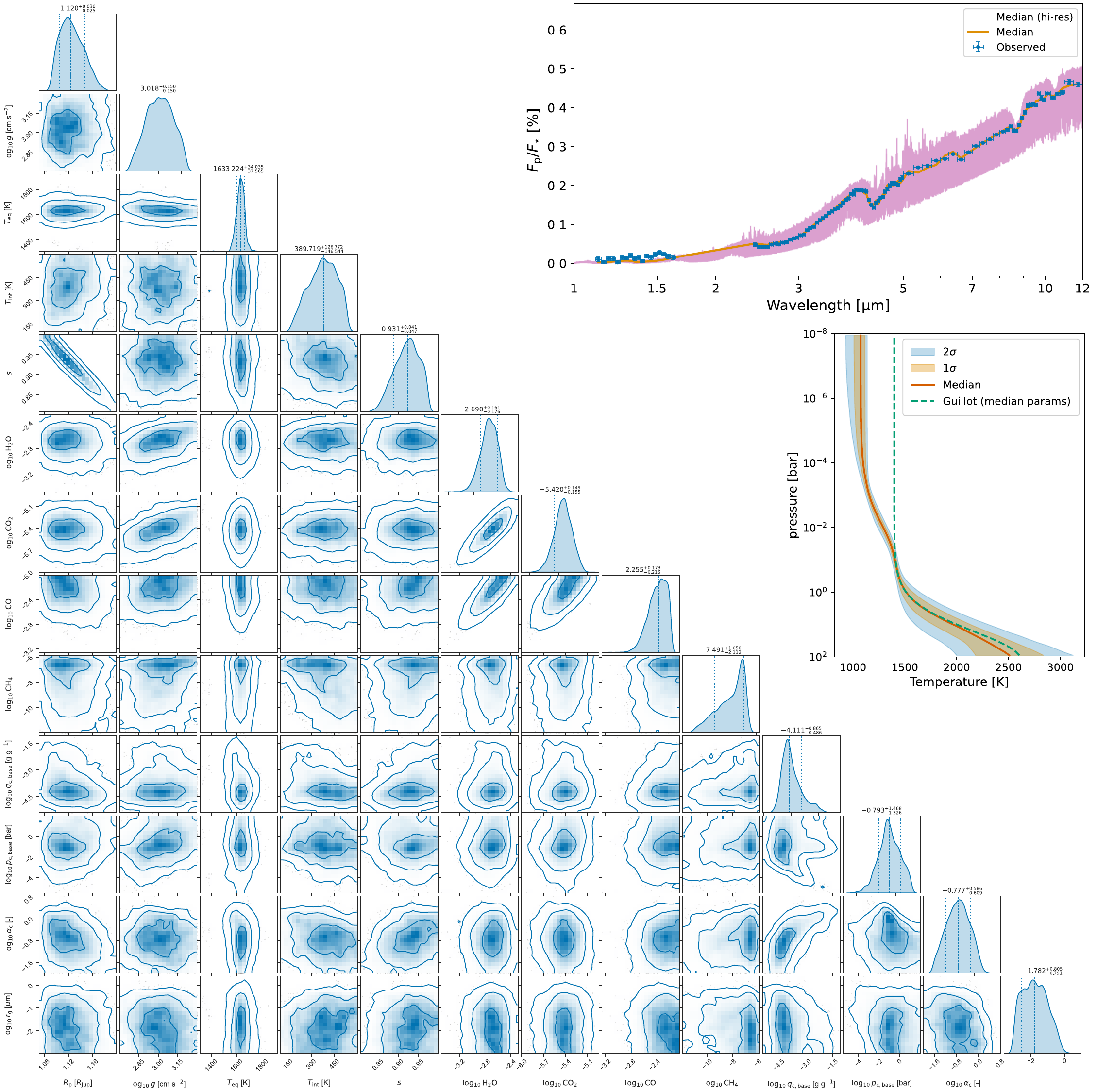}
    \caption{Equal weight posterior samples and kernel density estimation for the HD 189733b HST and JWST dayside emission spectra retrieval.
    The median and 1$\sigma$ confidence intervals are shown as the dashed and dotted lines respectively.
    The plots in the top right show the best fit median spectra (orange) with 1$\sigma$ (shaded) region to the observed transmission spectra data (blue), taken from \citet{Changeat_2022, Zhang_2025} and \citet{Inglis_2024} for the full spectral range (top right).
    The high-resolution median spectrum is in purple.
    The middle right plot shows the retrieved modified Guillot $T$-$p$ profile (Sect. \ref{sec:Guillot}), compared to the original formulation (green dashed line).}
    \label{fig:HD189b}
\end{figure*}

\begin{figure}
    \centering
    \includegraphics[width=\linewidth]{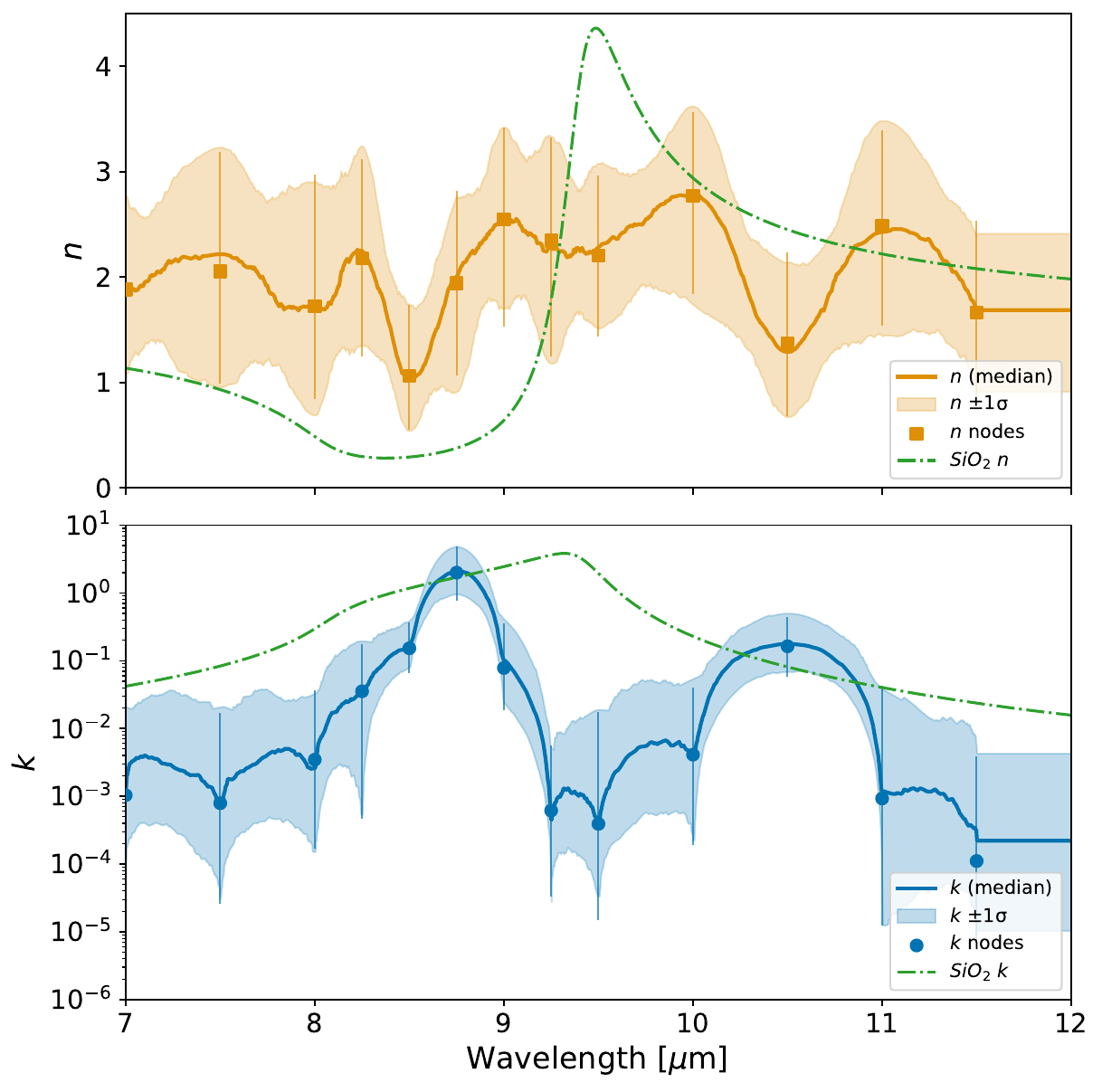}
    \caption{Direct $n$ (top orange) and $k$ (bottom blue) retrieval results from the HD 189733b emission spectrum test.
    The glassy $\alpha$-SiO$_{2}$ optical constants from \citet{Zeidler_2013} are plotted with the dash-dot green lines.}
    \label{fig:HD189b_nk}
\end{figure}

In this test, I perform a dayside emission spectra retrieval test on the canonical hot Jupiter HD 189733b.
I use a combined HST WFC3 \citep{Changeat_2022}, JWST NIRCam \citep{Zhang_2025} and JWST MIRI LRS \citep{Inglis_2024} dayside emission spectrum.
I use the direct ($n$, $k$) retrieval method in the MIRI wavelength regime, to try to recover the optical constants of the suspected aerosol absorption feature seen in \citet{Inglis_2024}.
Using \textit{dynesty}, the retrieval had 44 sampled parameters (Table \ref{tab:HD189b}), with the elapsed time from start to finish being 1 day 13 hours 31 minutes (OS at $R$ = 20,000).

Figure \ref{fig:HD189b} presents the posterior distribution and best fit median models compared to the observational data.
I recover a generally metallicity enhanced atmosphere, with H$_{2}$O, CO and CO$_{2}$ VMRs around 5-10$\times$ Solar, in line with previous retrieval modelling \citep{Zhang_2025}.
Overall, the JWST NIRCam and MIRI data are well fit by the median model, with the HST WFC3 less well fit.
This suggests potential multi-dimensional effects may be present, such as a hotspot that would increase near-IR fluxes but not significantly influence redder wavelengths.
Alternatively, a significant near-IR albedo from high altitude clouds would also raise the near-IR spectral flux \citep{Evans_2013}.
Multiple $T$-$p$ profile techniques or hotspot emulation \citep[e.g.][]{Feng_2016, Schlawin_2024, Wiser_2015} may therefore be appropriate to apply to this planet, but this is outside the scope of this example study.
However, a uniform data reduction between the HST and JWST data would need to be carried out to make a fair comparison.

Figure \ref{fig:HD189b_nk} shows the recovered ($n$, $k$) optical constants from the retrieval model.
Similar to the WASP-17b retrieval, I find that the $n$ constant is unconstrained, with large uncertainty across the wavelength regime.
However, for the $k$ constant, two main features are present: one bluer than the SiO$_{2}$ absorption peak \citep{Zeidler_2013} data and a redder peak at $\approx$10.5 $\mu$m.
A possibility is that the bluer peak is caused by a SiO$_{2}$ polymorph different to quartz as seen in the \citet{Moran_2024} study, who investigated the WASP-17b MIRI cloud absorption feature.
The redder peak is harder to compare to other species, but may be an amorphous silicate material, as those materials generally exhibit a redder peak compared to quartz \citep[e.g.][]{Wakeford_2015, Kitzmann_2018, Luna_2021, Mullens_2024}.
Overall, this suggests that the commonly assumed $\alpha$-SiO$_{2}$ quartz material may be an inaccurate assumption for this exoplanet.
I recover a median cloud base mass mixing ratio of $q_{\rm c, base}$ $\approx$ 10$^{-4.1}$, which is in line with around a 5-10x Solar enhancement of Si \citep{Asplund_2021}.

\subsection{Emission - Gliese 229 B}
\label{sec:G229B}

\begin{figure*}
    \centering
    \includegraphics[width=\linewidth]{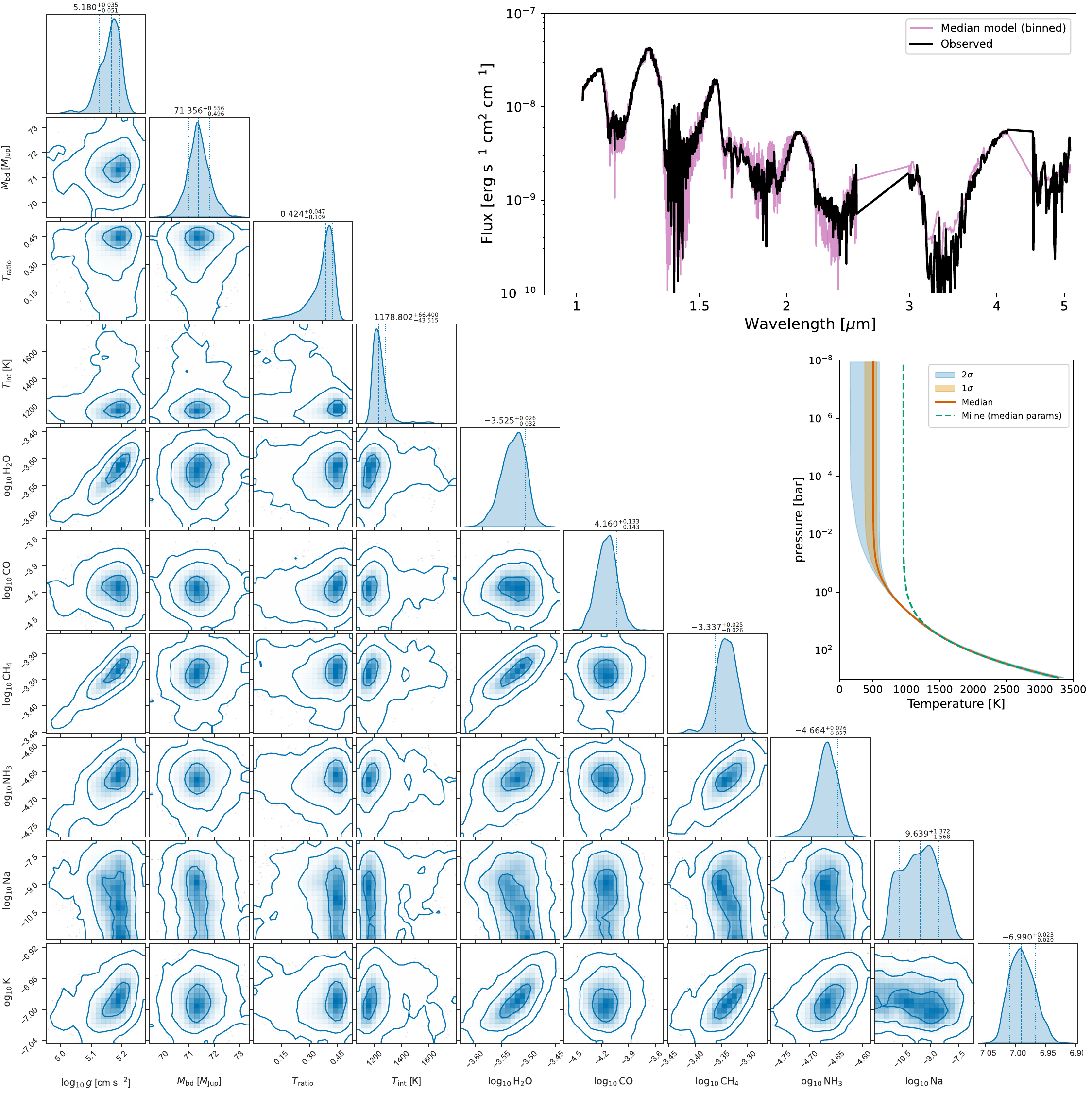}
    \caption{Equal weight posterior samples and kernel density estimation for the Gliese 229 B brown dwarf retrieval.
    The median and 1$\sigma$ confidence intervals are shown as the dashed and dotted lines respectively.
    The observational data (black) and median best-fit (purple) is shown in the top right panel.
    The median $T$-$p$ profile with 2$\sigma$ confidence interval is shown on the right middle panel.}
    \label{fig:G229B}
\end{figure*}

\begin{table}[]
    \centering
    \caption{Gliese 229 B brown dwarf emission spectrum retrieval setup and results.}    
    \begin{tabular}{c|c|c} \hline \hline
        Parameter & Prior  & Posterior \\ \hline
         $p_{\rm bot}$ [bar]&  $\delta$(1000) & - \\
         $p_{\rm top}$ [bar]&  $\delta$(10$^{-8}$) & - \\
         $p_{\rm ref}$ [bar]&   $\delta$(10)  & - \\
         $D$ [pc] &   $\delta$(10)  & - \\
         $\log_{10} g$ [cm s$^{-2}$]&  $\mathcal{U}$(3, 5.5) &  5.18$^{+0.03}_{-0.05}$\\
         $M_{\rm bd}$ [$M_{\rm jup}$] &  $\mathcal{N}$(71.4, 0.6) & 71.36$^{+0.54}_{-0.49}$ \\
         $T_{\rm int}$ [K] &  $\mathcal{U}$(500, 2000) &  1178$^{+63}_{-43}$\\
         $T_{\rm ratio}$ [-] &  $\mathcal{U}$(0, 1) & 0.43$^{+0.05}_{-0.11}$ \\
         $s$ [-] & $\mathcal{U}$(0, 1) & 0.92$^{+0.05}_{-0.07}$ \\
         $\log_{10} \kappa_{\rm ir}$ [cm$^{2}$ g$^{-1}$] &  $\mathcal{U}$(-6, 2) & -1.87$^{+0.07}_{-0.11}$  \\
         $\log_{10} p_{\rm t}$ [bar] &  $\mathcal{U}$(-8, 3) & 0.94$^{+0.14}_{-0.11}$ \\   
         $\beta$ [-] &  $\mathcal{U}$(0, 1) & 0.92$^{+0.06}_{-0.08}$\\
         $\log_{10}$ H$_{2}$O [-]& $\mathcal{U}$(-12, -1)  &  -3.52$^{+0.03}_{-0.03}$ \\
         $\log_{10}$ CO [-]& $\mathcal{U}$(-12, -1)  & -4.16$^{+0.14}_{-0.14}$ \\
         $\log_{10}$ CH$_{4}$ [-]& $\mathcal{U}$(-12, -1)  &  -3.34$^{+0.02}_{-0.03}$ \\ 
         $\log_{10}$ NH$_{3}$ [-]& $\mathcal{U}$(-12, -1)  & -4.66$^{+0.03}_{-0.03}$ \\ 
         $\log_{10}$ Na [-]& $\mathcal{U}$(-12, -1)  &  -9.61$^{+1.39}_{-1.56}$ \\
         $\log_{10}$ K [-]& $\mathcal{U}$(-12, -1)  &  -6.99$^{+0.02}_{-0.02}$ \\
         c [-]& $\mathcal{U}$(-15.63, -6.66)  &  -8.90$^{+0.01}_{-0.01}$ \\
         \hline \hline
    \end{tabular} 
    \label{tab:G229B}
\end{table}

To exhibit the brown dwarf retrieval system in \textit{Exo Skryer}, I perform a retrieval on the T-dwarf Gliese 229B, following the setup and collated observational data from \citet{Calamari_2022}.
\citet{Calamari_2022} used the \textit{Brewster} brown dwarf retrieval code presented in \citet{Burningham_2017}.
\citet{Calamari_2022} used data from \citet{Geballe_1996, Noll_1997} and \citet{Oppenheimer_1998} in their retrieval scheme.
Despite Gliese 229B being recently confirmed to be a binary system \citep{Brandt_2021, Xuan_2024, Whitebook_2024}, which will affect the interpretation of any retrieval results that assume a single object, the quality of the combined object spectral data and numerous previous modelling efforts on this system provide a useful comparison benchmark.
I perform a cloud-free, `mass constrained' retrieval following \citet{Calamari_2022}, taking as a prior, the dynamical mass measured by \citet{Brandt_2021}.
Using \textit{dynesty}, with 15 sampled parameters (Table \ref{tab:G229B}), the retrieval took 9 hours 8 minutes (OS at $R$ = 40,000).

Comparing to the results of \citet{Calamari_2022}, I retrieve reasonably similar VMRs (to within 1-2$\sigma$ typically) for all gas species as well as $T$-$p$ profile, suggesting both retrieval outcomes are approximately consistent with respect to the variables sensitive to radiative-transfer outcomes.
The main difference is that I recover a larger CO VMR; 10$^{-4.16}$ compared to 10$^{-4.56}$ in \citet{Calamari_2022}.
I recover a median radius of $R_{\rm bd}$ = 1.082 $R_{\rm jup}$.
The integrated spectrum of the median model resulted in $T_{\rm eff}$ = 909 K, which is around 80 K hotter than that retrieved in \citet{Calamari_2022}.
Utilising JWST MIRI \citep{Xuan_2024b} and future JWST NIRSpec (GO \#3762 P.I. J. Xuan) data in conjunction with current data should provide much needed clarity on the combined and individual parameters of each brown dwarf.

I found that using a free prior for mass $M_{\rm bd}$ $\mathcal{U}$(1, 80), \textit{Exo Skryer} exhibited the well-known Gravity-Radius-Composition degeneracy present in brown dwarf fitting and retrieval modelling of spectra \citep[e.g.][]{Burrows_2001, Zalesky_2019, Kitzmann_2020, Lueber_2022}.
The retrieval results consistently pushed against the upper prior boundary for the mass, producing unphysical parameter sets for the brown dwarf radius, for example, $R_{\rm bd}$ $<$  0.9 $R_{\rm jup}$.
The retrieval results were more consistent with the \citet{Kawashima_2025} study, who used ground-based high-resolution spectral data from Subaru/IRD and applied the \textit{ExoJAX} retrieval model \citep{Kawahara_2022, Kawahara_2025}.
They retrieved a larger mass, $\sim$100 $M_{\rm jup}$, in their free retrievals, and discussed the degeneracies encountered as applied to high-resolution spectra in detail.

\section{Discussion}
\label{sec:disc}

Through the numerous test retrievals presented in this study, \textit{Exo Skryer} has been shown to produce consistent and compatible retrieval outcomes to a wide variety of different frameworks, from \textit{NEMESIS} \citep{Irwin_2008, Barstow_2014}, \textit{petitRADTRANS} \citep{Molliere_2019}, \textit{POSEIDON} \citep{MacDonald_2023} and \textit{Brewster} \citep{Burningham_2017}.
Though \textit{Exo Skryer} does not yet have all the features present in the above cited frameworks, in particular currently lacking a chemical equilibrium solver, I have developed several new methods for \textit{Exo Skryer} that may be of use in other models, namely the modified Milne and Guillot $T$-$p$ profiles, non-grey cloud opacity methodology, and direct $n$ and $k$ optical constant retrieval.

In Section \ref{sec:direct_nk}, I proposed a new direct ($n$, $k$) optical constant retrieval method.
I found that the recovered optical constants are approximately consistent with the commonly assumed SiO$_{2}$ material composition, but the main peak is bluer for both the WASP-17b and HD 189733b data, suggesting that the optical constants for the $\alpha$-SiO$_{2}$ mineral may not be representative of the real cloud materials inside these hot atmospheres.
In addition, for the HD 189733b dayside emission case, I find evidence for a possible secondary absorption feature at around 10.5 $\mu$m.
\citet{Moran_2024} collated or derived optical constants for several SiO$_{2}$ polymorphs, showing that the specific formation conditions of the material are important in setting the shapes and peaks of the optical constants.
However, as noted in \citet{Moran_2024}, this data was patchy and possibly unreliable, warranting new laboratory experiments that produce exoplanet cloud samples at representative temperature conditions of these atmospheres that explore different crystalline states and potential mineral mixtures.

For the HD 189733b case, I intentionally only targeted the observed absorption region between 7-12 $\mu$m, ignoring the effects of cloud opacity on the rest of the spectrum.
This is due to the fact that, in emission, the scattering of cloud particles is critical to shaping the spectrum \citep[e.g.][]{Taylor_2021}, and the current scheme, using \citet{Kitzmann_2018}, does not parametrise the single scattering albedo or asymmetry parameter.
Therefore, the direct $n$-$k$ scheme should be utilised with caution for emission retrievals in its current form.
In addition, the silicate absorption feature may extend beyond 12 $\mu$m \citep[e.g.][]{Luna_2021, Suarez_2023} in some atmospheres, requiring further adjustment to the scheme.

A more complete retrieval methodology may be developed in the future, able to both retrieve optical constants at specific wavelengths and utilise a more general cloud scheme across the wavelength range.
In testing, I first attempted a Kramers–Kronig (KK) approach, retrieving the $k$ constant and using KK to calculate the $n$, similar to \citet{Irwin_2018}.
However, I found the computation to be too inefficient, and the retrieval unable to converge on a reasonable timescale.
An alternative may be to post-process the retrieved $k$ values using KK to produce physically consistent $n$ values that can be compared to current databases.
This method would also reduce the required number of retrieved parameters through ignoring the real component of the optical constants.
This is left to future iterations of the model.

In this study, I used an Nvidia GeForce RTX 4090 GPU for all retrievals which is sub-optimal for several reasons:
\begin{itemize}
    \item Double precision, 64 bit operations, critical for accurate forward model operation, are not optimised in the 4090 consumer grade architecture compared to server-grade equipment, reducing the numerical operation speed significantly.
    \item The 4090 was released in 2022, and is considered last generation hardware. 
    Newer GPUs may be even up to 10-100\% faster.
\end{itemize}
Therefore, I expect the specific values quoted in this study can be improved upon, possibly 2-5$\times$ using an optimised hardware setup. 

The speed tests in Section \ref{sec:speed} show that using \textit{pymultinest} is around two times faster than \textit{dynesty} for the 11 parameter HD 209458b case study.
Should expediency at lower dimensionality be required, \textit{pymultinest} offers the fastest retrieval option presented in this study.
For significantly larger dimensionality ($\sim$50+), other frameworks have been shown to be more efficient and robust \citep[e.g.][]{Handley_2015, Speagle_2020}.
Extremely parallel nested sampling frameworks such as \textit{blackjax-ns} \citep{Yallup_2025} are also being developed and have been applied to black hole merger events \citep{Prathaban_2025}.
This method fully parallelises over live points as well as likelihood calculations.
This is an active area of research, but shows promise to be the fastest sampling framework in the near future.

\section{Conclusion} 
\label{sec:conc}

With worldwide high-performance computing infrastructure continuing to hybridise to include a substantial GPU complement, computational demanding scientific codes such as exoplanet and brown dwarf retrieval models are shifting towards making efficient use of this new infrastructure landscape.
An added advantage of using highly-parallel models is that runtimes are significantly reduced compared to serial implementations, allowing faster exploration of Bayesian evidence and parameter estimation in the wide-wavelength, medium spectral resolution JWST era.

I present a new sub-stellar atmosphere retrieval model, \textit{Exo Skryer}, that leverages the JAX library for Python.
The design of \textit{Exo Skryer} allows scalable computation on CPU and GPU systems, as well as a mixture of the two, where in this study I found that the hybrid CPU+GPU option to be optimal, enabling robust, well tested sampling frameworks to be used while heavily accelerating forward model evaluations.
This significantly reduces the time for retrieval completion, down to minutes for simple models and handfuls of hours depending on the complexity of the forward model and dimensionality of the problem.
The added computational efficiency of \textit{Exo Skryer} enables heavier, many parameter, forward models or more computationally demanding physics and chemistry modules to be utilised without increasing the runtime beyond unfeasibility.
Large scale model comparison was not performed in this study, but is foreseen to also be accelerated.
Through using a JAX framework, \textit{Exo Skryer} is well placed to integrate current and future machine learning techniques, novel highly parallel sampling methodologies, and to take advantage of technological developments in GPU hardware.

\textit{Exo Skryer} offers a new microphysical parameter based, cloud opacity retrieval methodology using the mass mixing ratio profile of the cloud as the basic unit of the cloud opacity scheme.
This allows retrieval of parameters that can be more directly physically interpreted from cloud microphysical theory.
In addition, I presented a new, simple, direct optical constant retrieval methodology for recovering ($n$, $k$) constants for suspected JWST MIRI aerosol absorption bands.
The retrieval results of WASP-17b and HD 189733b suggest that the commonly assumed SiO$_{2}$-quartz composition may not be fully representative of the real material in hot gas giant atmospheres.
Including advanced techniques such as leave-one-out (LOO) methodologies \citep[e.g.][]{Welbanks_2023} into \textit{Exo Skryer}, to test the sensitivity of Bayesian evidence to specific data points, would be beneficial.

Lastly, the retrievals performed in this study used older, non-optimal hardware, suggesting that further speed up of retrieval times than that presented here is possible.
This paper acts as the basic reference for \textit{Exo Skryer}, with many updates expected in the future as utilisation increases.
\textit{Exo Skryer} is available as open-source software on GitHub\footnote{\url{https://github.com/ELeeAstro/Exo_Skryer}}.

\section*{Acknowledgments}
I thank The Open Journal of Astrophysics for providing an accessible peer review outlet for larger, technical papers without publication fees, which are increasingly becoming out of reach for independent early career scientists.
I acknowledge funding from the CSH through the Bernoulli Fellowship.
I greatly thank B. Morris for discussions and advice on proper Python package development, JAX and sampling techniques, as well as help with setting up the \textit{Exo Skryer} online documentation drivers.
I thank J. Taylor for reading an initial draft of the paper, and providing advice on the retrieval results.
I thank D. Kitzmann, L. Welbanks and V. Parmentier for various discussions and advice on retrieval modelling particulars.
I thank D. Bower for encouragement and advocacy on the use of JAX for this project.

\bibliographystyle{mn2e} 
\bibliography{main_oja.bib} 

\end{document}